\documentclass[12pt,preprint]{aastex}
\def\apj{{ApJ}}
\def\mnras{{ MNRAS}}

\def\be{\begin{equation}}
\def\ee{\end{equation}}
\def\bea{\begin{eqnarray}}
\def\eea{\end{eqnarray}}

\usepackage{graphicx}
\usepackage{ulem}
\usepackage{subfigure}
\usepackage[colorlinks,linkcolor=blue,anchorcolor=red,citecolor=cyan]{hyperref}

\begin{document}

\title{Fast $\gamma$-ray variability in blazars beyond redsift 3}

\author{Shang Li\altaffilmark{1,2}, Zi-Qing Xia\altaffilmark{1}, Yun-Feng Liang\altaffilmark{1}, Neng-Hui Liao\altaffilmark{1}, Yi-Zhong Fan\altaffilmark{1}}
\altaffiltext{1}{Key Laboratory of Dark Matter and Space Astronomy, Purple Mountain Observatory, Chinese Academy of Sciences, Nanjing 210008, China}
\email{liaonh@pmo.ac.cn} \email{yzfan@pmo.ac.cn}
\altaffiltext{2}{University of Chinese Academy of Sciences, Yuquan Road 19, Beijing 100049, China}

%\thanks{E-mail:liaonh@pmo.ac.cn (NHL);yzfan@pmo.ac.cn (YZF)}

\begin{abstract}
High-redshift blazars are one of the most powerful monsters in the universe and $\gamma$-ray variability carries crucial information of their relativistic jets. In this work we present results of the first systematical temporal analysis of {\it Fermi}-LAT data of all known seven $\gamma$-ray blazars beyond redshift 3. Significant long-term $\gamma$-ray variability is found from five sources in monthly $\gamma$-ray light curves, in which three of them are reported for the first time. Furthermore, intraday $\gamma$-ray variations are detected from NVSS J053954$-$283956 and NVSS J080518$+$614423. Doubling variability timescale of the former source is limited as short as $\lesssim$ 1 hour (at the source frame). Together with variability amplitude over one order of magnitude, NVSS J053954$-$283956 is the most distant $\gamma$-ray flaring blazar so far. Meanwhile, intraday optical variability of NVSS J163547$+$362930 is found based on archival PTF/iPTF light curve. Benefited from multiwavelength activity of these sources, constraints of their Doppler factors as well as locations of $\gamma$-ray radiation region and indications for the SDSS high redshift jetted active galactic nuclei deficit are discussed.
\end{abstract}


\section{Introduction}
Blazars, including flat spectrum radio quasars (FSRQs) and BL Lacertae objects (BL Lacs), are radio-loud active galactic nuclei (AGNs) whose relativistic jets are closely pointing to our line of sight \citep{BR78,1995PASP..107..803U}. The jet viewing angle $\theta_{v}$ is smaller than or comparable to the jet beaming angle (1/$\Gamma$, where $\Gamma$ is the jet bulk Lorentz factor), hence the jet emission is strongly boosted because of relativistic effects, making blazars so luminous that they are visible even at very high redshifts \citep[e.g.,][]{2004ApJ...610L...9R,2014ApJ...795L..29Y}. High-redshift blazars are extremely powerful monsters harboring super-massive black holes (SMBHs) heavier than one billion solar masses \citep[e.g.,][]{2010MNRAS.405..387G}. These sources are of major importance since they shed lights on formation and growth of the first generation of SMBHs and the role that relativistic jets play at that time \citep[e.g.,][]{2010A&ARv..18..279V,2015aska.confE..71A}. Studies of high-redshift blazars also provide information on whether and how the jet properties change with cosmic time, along with their potential impact on evolution of AGNs as well as their host galaxies \citep[e.g.,][]{2012ARA&A..50..455F,2011MNRAS.416..216V,2013Sci...341.1082M}.

Since the relativistic jets are responsible for bright radiation of blazars, they are characterized by highly variable and polarized nonthermal continuum emissions, generally detected in all observable bands from radio to $\gamma$-ray (both GeV and TeV) regimes \citep{1997ARA&A..35..445U}. Their spectral energy distribution (SED) reveals a two-bump shape, where one is likely due to synchrotron emission of relativistic electrons in magnetic fields and the other extending to $\gamma$ rays is usually explained as inverse Compton scattering of soft photons by the same population of relativistic electrons \citep[e.g.,][]{1992ApJ...397L...5M,1993ApJ...416..458D,1994ApJ...421..153S,2000ApJ...545..107B}. One of the extraordinary phenomena in blazars is their fast $\gamma$-ray variability. Imaging atmospheric Cerenkov telescopes (IACTs), like HESS and MAGIC, have detected flux changes on a time scale of a few minutes in high frequency peaked BL Lacs PKS 2155$-$304 \citep{2007ApJ...664L..71A} and Mkn 501 \citep{2007ApJ...669..862A}. Similar extreme behaviors have been also detected in other subtypes of jetted AGNs, including the low frequency peaked BL Lac BL Lacertae \citep{2013ApJ...762...92A} and FSRQ PKS 1222+21 \citep{2011ApJ...730L...8A}, strongly suggesting it is a common feature independent on the source type. In GeV domain, flux variations on a time scale of several hours have been found in a few sources since the CGRO era \citep[e.g.,][]{1997ApJ...476..692M,1998ApJ...497..178W}. Number of such detections has significantly increased right now thanks to the latest generation of space $\gamma$-ray telescope {\it Fermi}, though the minimum variability timescale is generally limited to a few hours due to its routine survey observation mode (e.g., \citealt{2011A&A...530A..77F,2013ApJ...767..103V,2015NewA...34..134L,2016ApJS..226...17L}, but also see \citealt{2013A&A...555A.138F}). Recently, taking advantage of a target of opportunity (ToO) repointing of {\it Fermi} Large Area Telescope (LAT, \citealt{2009ApJ...697.1071A}), extremely fast GeV $\gamma$-ray flux variability with doubling time less than 5 minutes has been detected in a giant outburst of 3C 279 \citep{2016ApJ...824L..20A}. Fast $\gamma$-ray variability of blazars is important for investigating the speed, composition and energetics of relativistic jets \citep[e.g.,][]{2008MNRAS.384L..19B}.

Detected number of high redshift blazar candidates (i.e. $z$ $\gtrsim$ 4) has significantly increased over the last decade, thanks to unprecedentedly powerful X-ray observatories (e.g. {\it Swift}; \citealt{2004ApJ...611.1005G}) as well as wide-field surveys in the optical/UV (e.g. the Sloan Digital Sky Survey, SDSS; \citealt{2000AJ....120.1579Y}), infrared ({\it WISE}; \citealt{2010AJ....140.1868W}), and radio bands (e.g. the NRAO VLA Sky Survey, NVSS; \citealt{1998AJ....115.1693C}). So far the farthest known blazar candiate is Q0906+6930 ($z$ = 5.48; \citealt{2004ApJ...610L...9R}), along with several others beyond redshift 5, including SDSS J102623.61+254259.5 ($z$ = 5.2; \citealt{2012MNRAS.426L..91S}), SDSS J013127.34$-$032100.1 ($z$ = 5.18; \citealt{2014ApJ...795L..29Y}) and SDSS J114657.79+403708.6 ($z$ = 5.0; \citealt{2014MNRAS.440L.111G}). These sources are identified as blazars in consideration of their high radio loudness and hard X-ray spectra \citep[e.g.,][]{2006AJ....132.1959R}. In $\gamma$-ray perspective, all $\gamma$-ray AGNs beyond redshift 3 are FSRQs and PKS 0537$-$286 ($z$ = 3.1; \citealt{1978ApJ...226L..61W}) had stood as the farthest $\gamma$-ray blazar for a long time \citep{2010dApJ...715..429}. Recently, 5 new $\gamma$-ray blazars with redshifts over 3.3 have been identified by {\it Fermi}-LAT Collaboration based on spatially overlapping between the $\gamma$-ray sources and their optical counterparts as well as their typical $\gamma$-ray FSRQ shape SEDs (i.e. high Compton dominance; \citealt{2017ApJ...837L...5A}). However, variability information of these sources is still lacking.

Here we perform a detailed analysis of {\it Fermi}-LAT $\gamma$-ray data of all known 7 $\gamma$-ray blazars beyond redshift 3. This work is organized as follows: in section \ref{sec2}, strategies in the {\it Fermi}-LAT data analysis procedure is introduced; results of the analysis are reported in section \ref{sec3}, in which we mainly focus on the results in the temporal domain; finally, in section \ref{sec4} we summarize our results with some discussions.

\section{LAT DATA ANALYSES}
\label{sec2}

In this paper, we used publicly released {\it Fermi}-LAT {\tt Pass 8} data ({\tt P8R2\_SOURCE\_V6}, {\tt FRONT+BACK}) and updated {\it Fermi} {\tt Science-Tools} package of version {\tt v10r0p5} to perform the analysis. For each target, we considered the data set within a $10^{\circ}$ region of interest (ROI) from 2008 October 27 to 2017 June 12 (i.e., Mission Elapsed Time, MET, between 246823875 s and 518983875 s), along with energy range from 100 MeV to 100 GeV. In order to reduce contamination from the Earth's limb, we removed $\gamma$-ray events with zenith angle larger than $90^{\circ}$. The recommended quality-filter cuts (i.e. {\tt DATA\_QUAL==1 \&\& LAT\_CONFIG==1}) have been followed to ensure that the spacecraft keeps in a good condition and hence the data set is valid for science use.

Firstly, we used the script {\tt make3FGLxml.py}\footnote{\url{http://fermi.gsfc.nasa.gov/ssc/data/analysis/user/}} to generate an initial background model for each target, which includes all point source within $15^{\circ}$ in the third {\it Fermi}-LAT source catalog (3FGL) \citep{2015ApJS..218...23A} as well as the latest galactic diffuse $\gamma$-ray emission model {\tt gll\_iem\_v06.fits} and isotropic emission template {\tt iso\_P8R2\_SOURCE\_V6\_v06.txt}\footnote{\url{http://fermi.gsfc.nasa.gov/ssc/data/access/lat/BackgroundModels.html}} \citep{2016ApJS..223...26A}. $\gamma$-ray locations and spectral templates (i.e. power-law function, $dN/dE \propto E^{-\Gamma}$, where $\Gamma$ is the spectral index) of the targets were set same as that in the literature \citep{2015ApJS..218...23A,2017ApJ...837L...5A}. Then {\tt unbinned} likelihood algorithm implemented in the {\tt gtlike} task was adopted to extract the flux and spectrum. During the extraction, the flux and spectral parameters of sources within the $10^{\circ}$ ROI, together with normalization factors of the two diffuse components, were set free. Significance of the target is quantified with the test statistic (TS), defined as TS$=-2{\rm ln}({L_{0}/L}$) \citep{1996ApJ...461..396M}, where ${L}$ and $L_{0}$ are the maximum likelihood values for the model with and without target source, respectively. Since 3FGL is based on 4-year {\it Fermi}-LAT data that has narrower data time range than ours, we have checked whether there are newly emerging $\gamma$-ray sources beyond 3FGL by generating a residual TS map for each ROI. All new $\gamma$-ray background sources corresponding to excesses with TS $>25$ (i.e. detection significance of 4.2 $\sigma$) in the TS maps were added into the background model. For these new sources, their spectral models were also set to be power-law function and $\gamma$-ray positions were obtained by the {\tt gtfindsrc} task. After all these steps, the final results were obtained by analyzing the updated background model.

In temporal analysis, for background sources neither bright (i.e. TS value twice that of the target) nor close ($<$ 4$^\circ$) to the targets, their spectral indexes were frozen to values from the global fit. Meanwhile, considering high Galactic latitude background point sources that are dominated by blazars with highly variable $\gamma$-ray emissions, we removed the weak background sources (i.e. TS $<$ 1) from the source model. When a special $\gamma$-ray flare event was focused, firstly, an analysis of the entire flare epoch (i.e. several tens of days) was performed. Then short-term light curves were extracted to search evidences of intraday $\gamma$-ray variability. During the extraction, normalizations of two diffuse emission components and spectral parameters of all background sources were fixed to values from the analysis covering the whole flaring period. Note that this strategy has been also adopted for several similar studies \citep[e.g.,][]{2013ApJ...766L..11S,2015NewA...34..134L,2016ApJS..226...17L,2016ApJ...824L..20A}. Meanwhile, we also took into account the $\phi$ dependence of effect area of {\it Fermi}-LAT due to its square shape, and set {\tt phibins=5} in the {\tt gtltcube} task then. When TS value of the target is smaller than 4, {\tt pyLikelihood UpperLimits} tool was adopted to calculate the 95\% upper limit of the flux.

\section{RESULTS}
\label{sec3}
\subsection{Global analysis}
There are in total 7 known $\gamma$-ray blazars beyond redshift 3 so far. Their basic information are listed in Table ~\ref{table:tb1}. Two sources (i.e. NVSS J053954$-$283956 and NVSS J080518$+$614423) are known as relatively strong $\gamma$-ray sources and have been included in {\it Fermi} source catalogs \citep[e.g.,][]{2010ApJS..188..405A,2015ApJS..218...23A}. $\gamma$-ray emissions of other five sources have been detected from a recent search of high redshift $\gamma$-ray blazars performed by {\it Fermi} collaboration \citep{2017ApJ...837L...5A}. Redshift range of these 7 sources is between 3.0 and 4.3, especially NVSS J151002$+$570243 locates beyond redshift 4. Except NVSS J053954$-$283956 whose SMBH mass listed here is obtained from modeling its big blue bump from the accretion disk \citep{2010MNRAS.405..387G}, other estimations are based on optical spectroscopic observations \citep{2012RMxAA..48....9T,2015ApJS..219...12A}.

Before performing the temporal analysis, a fit of the entire $\sim105$ months LAT data for each source has been accomplished. All 7 sources are found as significant $\gamma$-ray emitters with soft spectra (i.e. $\Gamma_{\gamma}>$2.7), which is consistent with the literature \citep{2015ApJS..218...23A,2017ApJ...837L...5A}. Averaged $\gamma$-ray fluxes as well as corresponding $\gamma$-ray spectral indexes and TS values are also provided in Table ~\ref{table:tb1}. NVSS J053954$-$283956 is the brightest one among these sources, along with the highest TS value. Note that for sources selected from \cite{2017ApJ...837L...5A}, our averaged fluxes are generally lower than their values, because we chose a different energy range of LAT data (i.e. 0.1 GeV $-$ 100 GeV) from theirs (i.e. 0.06 GeV $-$ 300 GeV). For sources with TS $>$ 100, sophisticated spectral model (i.e. Log-parabolic function) is adopted to fit the entire LAT data, but no significant improvement is found compared to the initial usage of powerlaw function.

\subsection{Temporal behaviors}
\label{sec:temporal}
\subsubsection{Monthly $\gamma$-ray light curves}
\label{sec:longterm}
\begin{figure}
\centering
\includegraphics[width=0.4\textwidth]{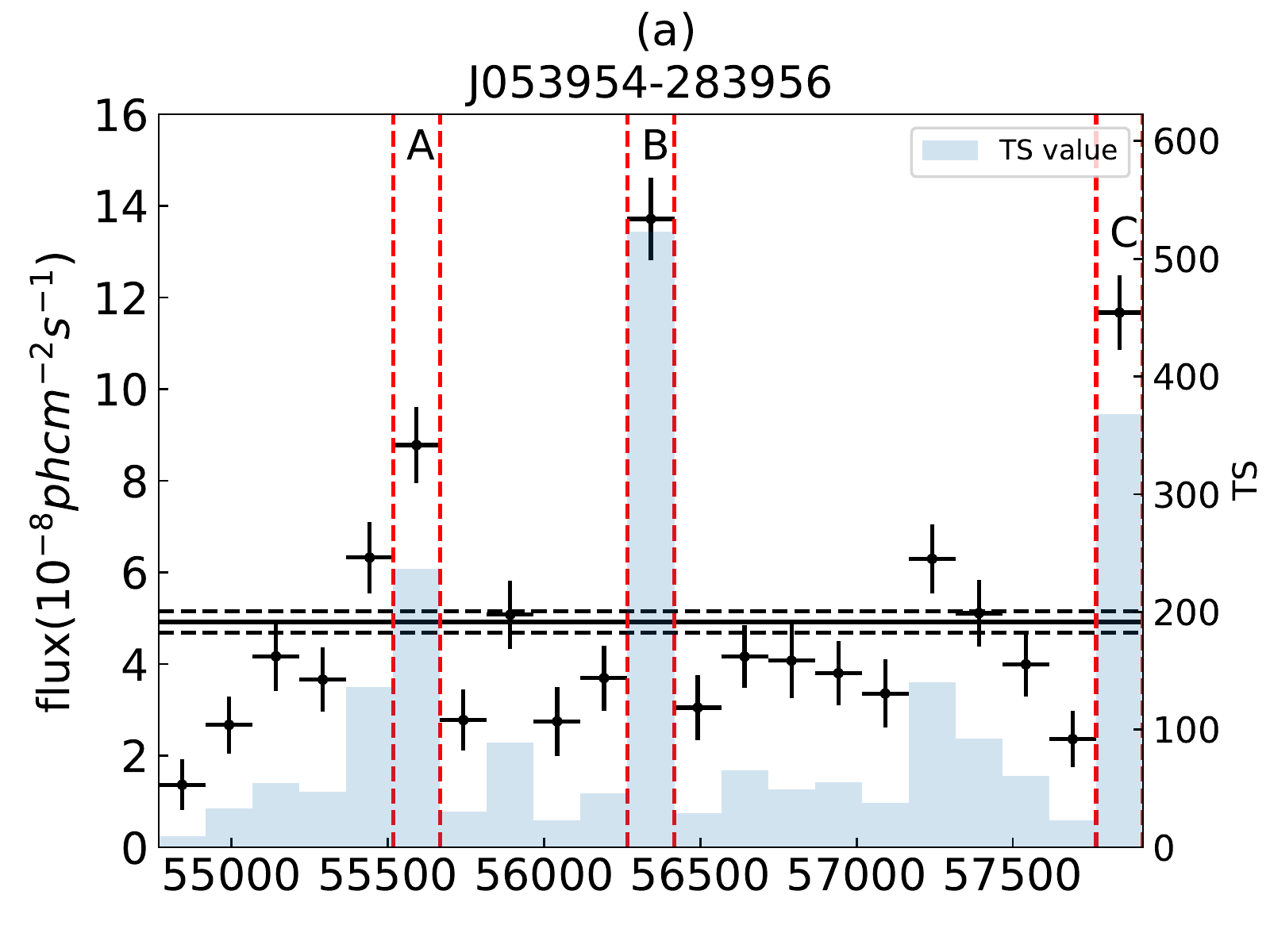}
\includegraphics[width=0.4\textwidth]{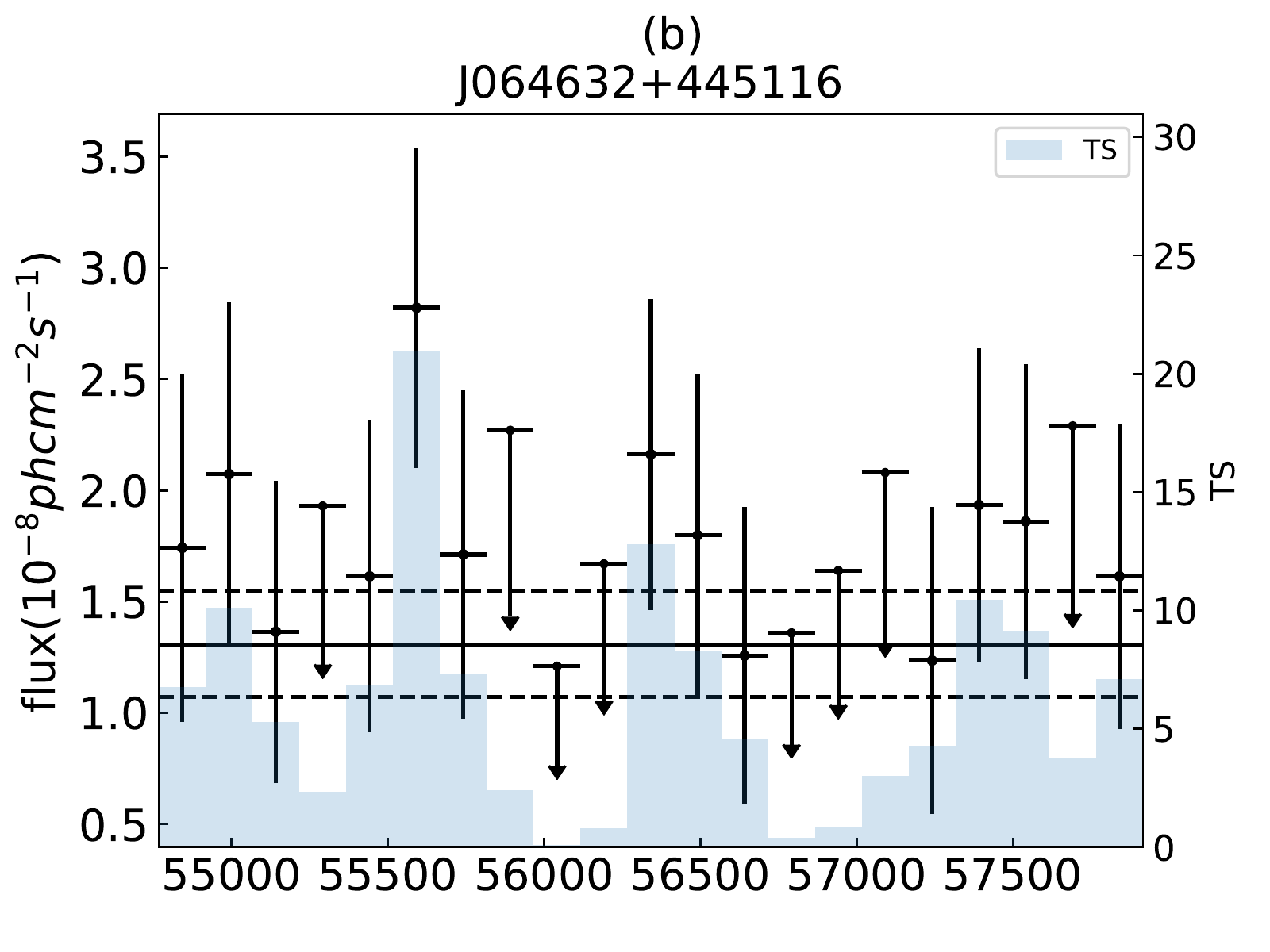}
\includegraphics[width=0.4\textwidth]{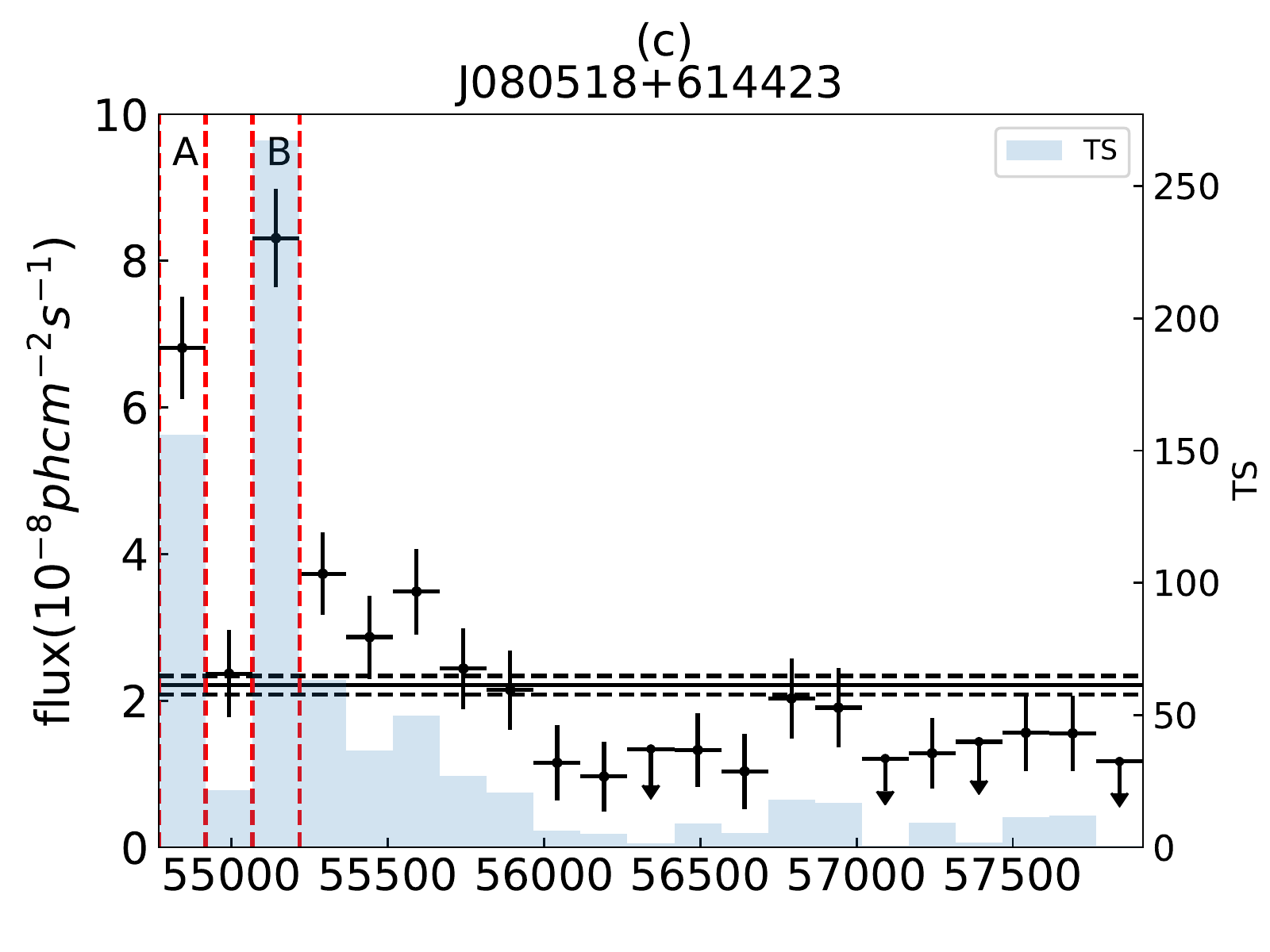}
\includegraphics[width=0.4\textwidth]{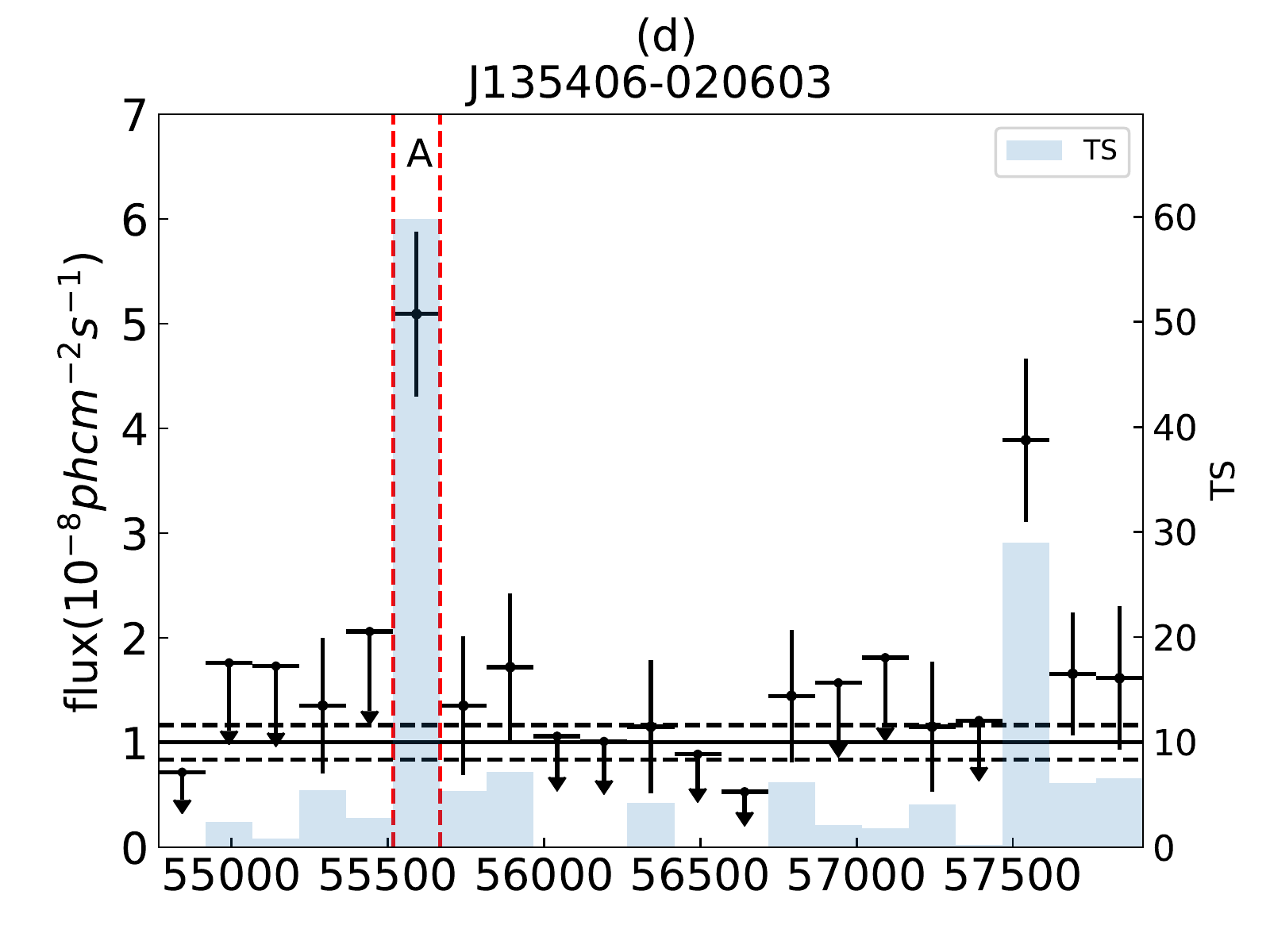}
\includegraphics[width=0.4\textwidth]{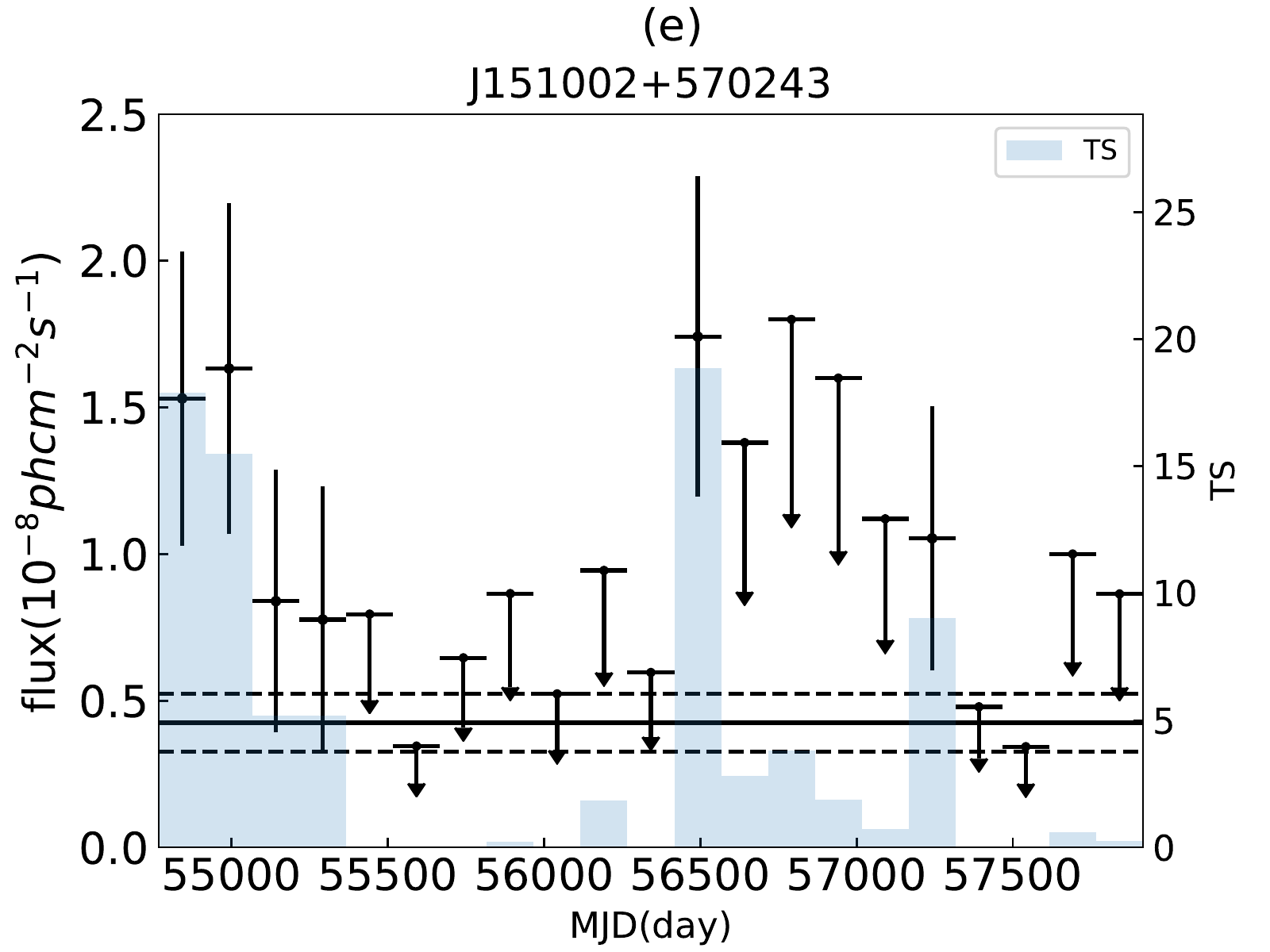}
\includegraphics[width=0.4\textwidth]{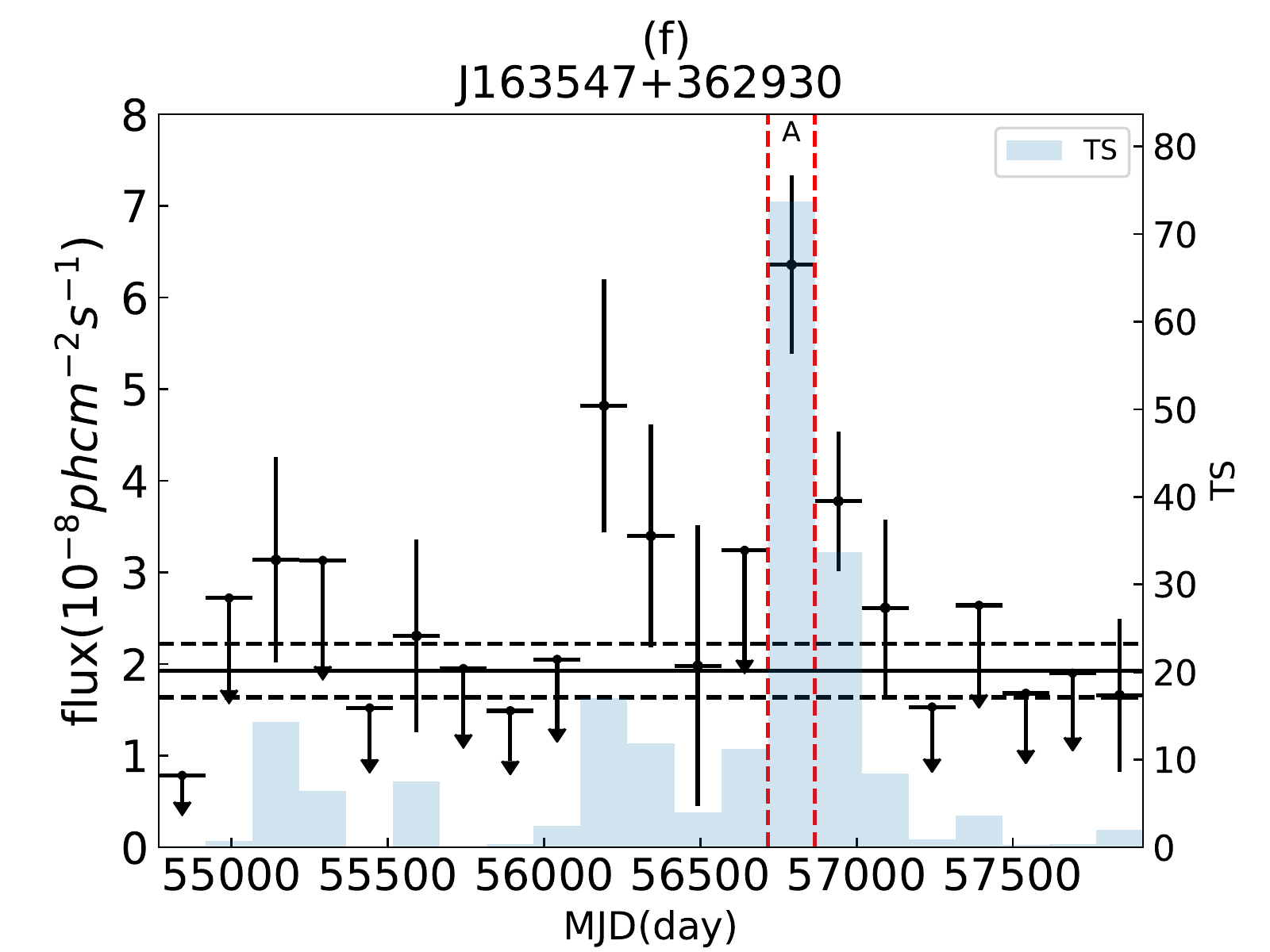}
\includegraphics[width=0.4\textwidth]{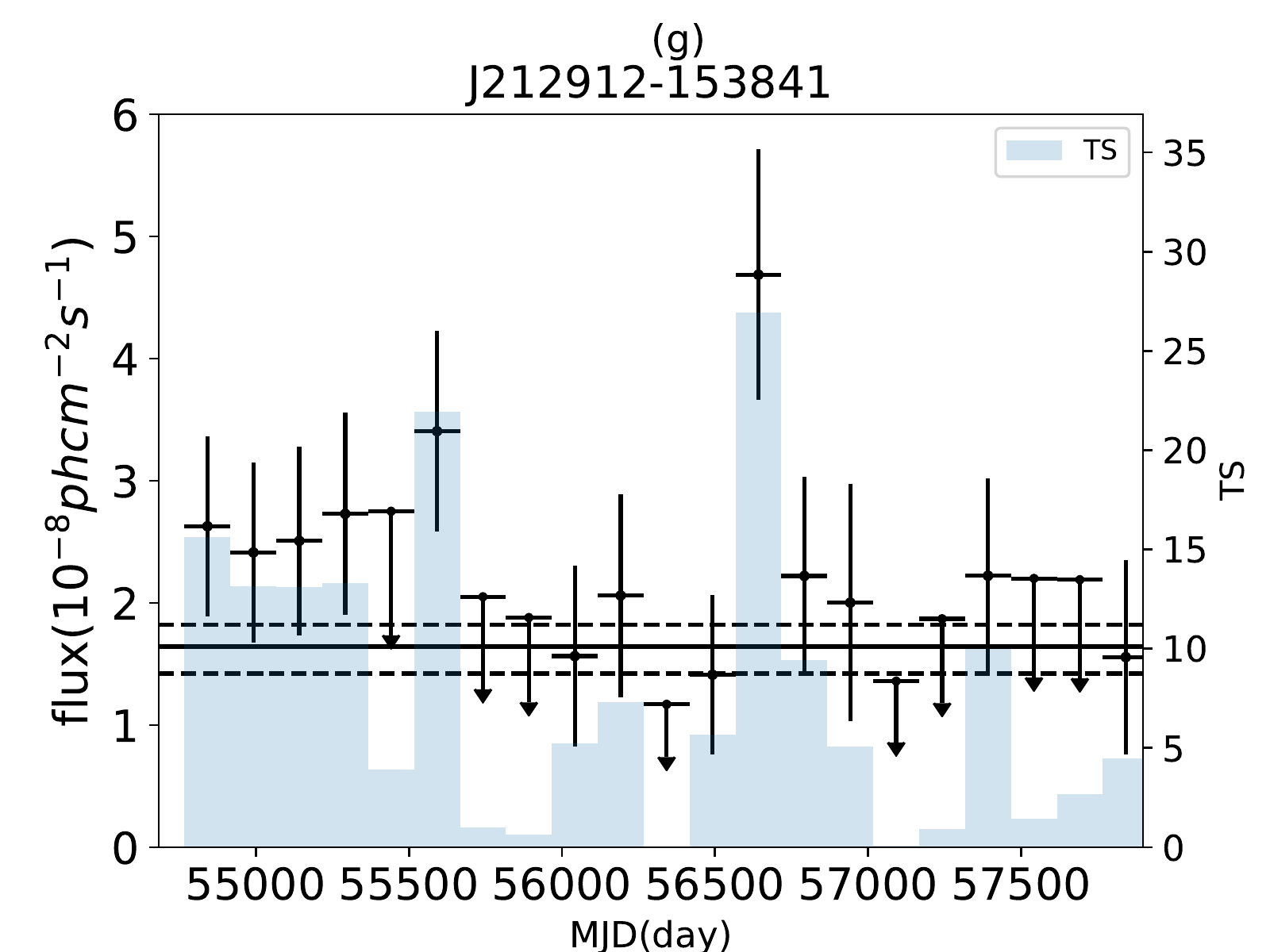}
\caption{100 MeV - 100 GeV monthly light curves of 7 high-reshift $\gamma$-ray blazars. Horizontal solid line along with two dashed lines in each panel represent the average flux and its $1\sigma$ error range derived in the global analysis, respectively. Red dashed vertical lines represent time epochs when further temporal analyses are performed. If any significant $\gamma$-ray flares appear, they are marked as different capitals.}
\label{fig:fiveLC}
\end{figure}

\begin{figure}
\centering
\includegraphics[width=0.8\textwidth]{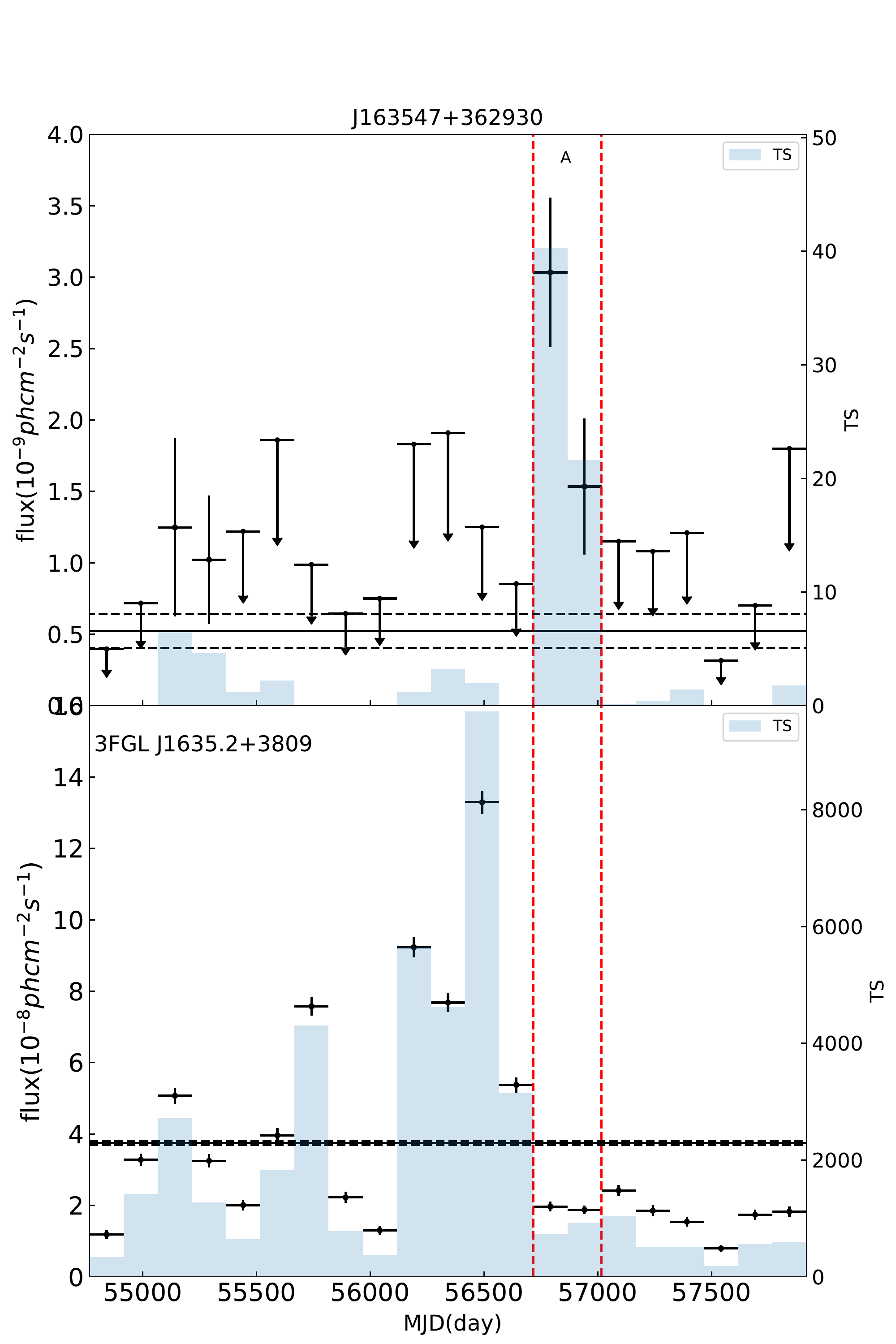}
\caption{500~MeV - 100~GeV monthly light curve of NVSS J163547+362930 as well as its neighbor 3FGL J1635.2$+$3809. Horizontal solid line and two dashed lines in each panel represent the average flux and its $1\sigma$ error range derived in the global analysis, respectively. Red dashed vertical lines represent the time epoch when the 15-day bin light curves are extracted.}
\label{fig:500mev}
\end{figure}

Since majority of our sources are not bright in $\gamma$ rays (i.e. $\rm TS_{105~month} \lesssim$ 100), firstly, we evenly divide the total LAT data into 21 time bins (with each bin about 5 months) to extract a $\gamma$-ray light curve for each source, see Figure \ref{fig:fiveLC}. For NVSS J053954$-$283956 and NVSS J080518$+$614423, significant $\gamma$-ray variability is obvious, confirming the results in {\it Fermi} source catalogs \citep[e.g.,][]{2012ApJS..199...31N}. While for other five sources, their error bars are relatively large and there are many upper limits in the light curves. So we use ``variability index'' test \citep{2012ApJS..199...31N} to quantify significance of variability of the light curves, from which information of the upper limits can be properly considered. The null hypothesis of this test is that source flux is constant across the full time period. The variability index is derived using following expression in \cite{2012ApJS..199...31N},
\begin{equation}
{\rm TS_{var}}=-2\,\sum_{i} \frac{\Delta F_{i}^2}{\Delta F_{i}^2+f^2 F_{\rm const}^2} {\rm ln}\left(\frac{\mathcal{L}_{i}(F_{\rm const})}{\mathcal{L}_{i}(F_{i})}\right),
\label{eq:tsvar}
\end{equation}
where for the $i$-th time bin, $\mathcal{L}_{i}$ is the likelihood value, $F_{i}$ is the observed photon flux, $\Delta F_{i}$ is the statistical uncertainty of $F_{i}$, $F_{\rm const}$ is the assumed constant flux, and $f$ is the systematic correction factor which we take a value of 2\% following \cite{2012ApJS..199...31N}. In our analysis, the optimized constant flux for each source is close to the average flux from the global analysis (within 1$\sigma$ statistical uncertainty). If the null hypothesis is correct, the derived $\rm TS_{var}$ in case of a light curve with 21 time bins should follow a $\chi^{2}$ distribution with 20 degrees of freedom and hence the variability significance in $\sigma$ unit is obtained (also listed in Table \ref{table:tb1}). It is not surprised that $\sigma_{var}$ values of NVSS J053954$-$283956 and NVSS J080518$+$614423 are high (i.e.$>$ 10). Interestingly, we find $\gamma$-ray emissions of NVSS J135406$-$020603, NVSS J151002$+$570243 and NVSS J163547$+$362930 are significantly variable (i.e. $\sigma_{var}>$ 3). Due to the limited angular resolution of {\it Fermi}-LAT strong variability in nearby background sources can cause artificial variability for the target. Therefore, we have checked whether there are any such neighbors around these five sources. We find that this situation only happens to NVSS J163547$+$362930, for which there is one bright and highly variable background source 3FGL J1635.2$+$3809 about $1.6^{\circ}$ away. Since the 68\% C.L. contamination angle of LAT for 500~MeV photons is about $1.5^{\circ}$, to avoid significant impact from the neighbor, individual light curves with lower energy cut of LAT data raising from 100~MeV to 500~MeV both for the target and the neighbor are extracted (see Figure \ref{fig:500mev}). There are two major $\gamma$-ray flares in the $>$ 100~MeV light curve of NVSS J163547$+$362930, with peaking time around MJD 56192 and MJD 56792, respectively. The former flare coincides with a high flux state of the neighbor 3FGL J1635.2$+$3809 and disappears in the $>$ 500~MeV light curve, indicating that it is probably artificial. However, the other flare corresponds to a low flux state of the neighbor and remains to be significant in the $>$ 500~MeV light curve, which suggests an intrinsic link between this flare and the target. Variability index of the $>$ 500~MeV light curve of NVSS J163547$+$362930 is also calculated, given as $\sigma_{var}=$ 4.9. Although it is smaller than that in $>$ 100~MeV case, the $\gamma$-ray emission of NVSS J163547$+$362930 is still proved to be significantly variable.

\subsubsection{Detecting fast $\gamma$-ray variability}
According to the monthly $\gamma$-ray light curves, fluxes of several time bins are significantly higher than the averaged flux. Together with relatively large TS values (i.e. $>50$), it allows us to further search for any possible fast $\gamma$-ray variability. Except for NVSS J135406-020603 that no significant variability in short-term is found, detailed temporal analyses of NVSS J053954$-$283956, NVSS J080518+614423 and NVSS J163547+362930 are described below.

\begin{figure}
\centering
\includegraphics[width=0.42\columnwidth]{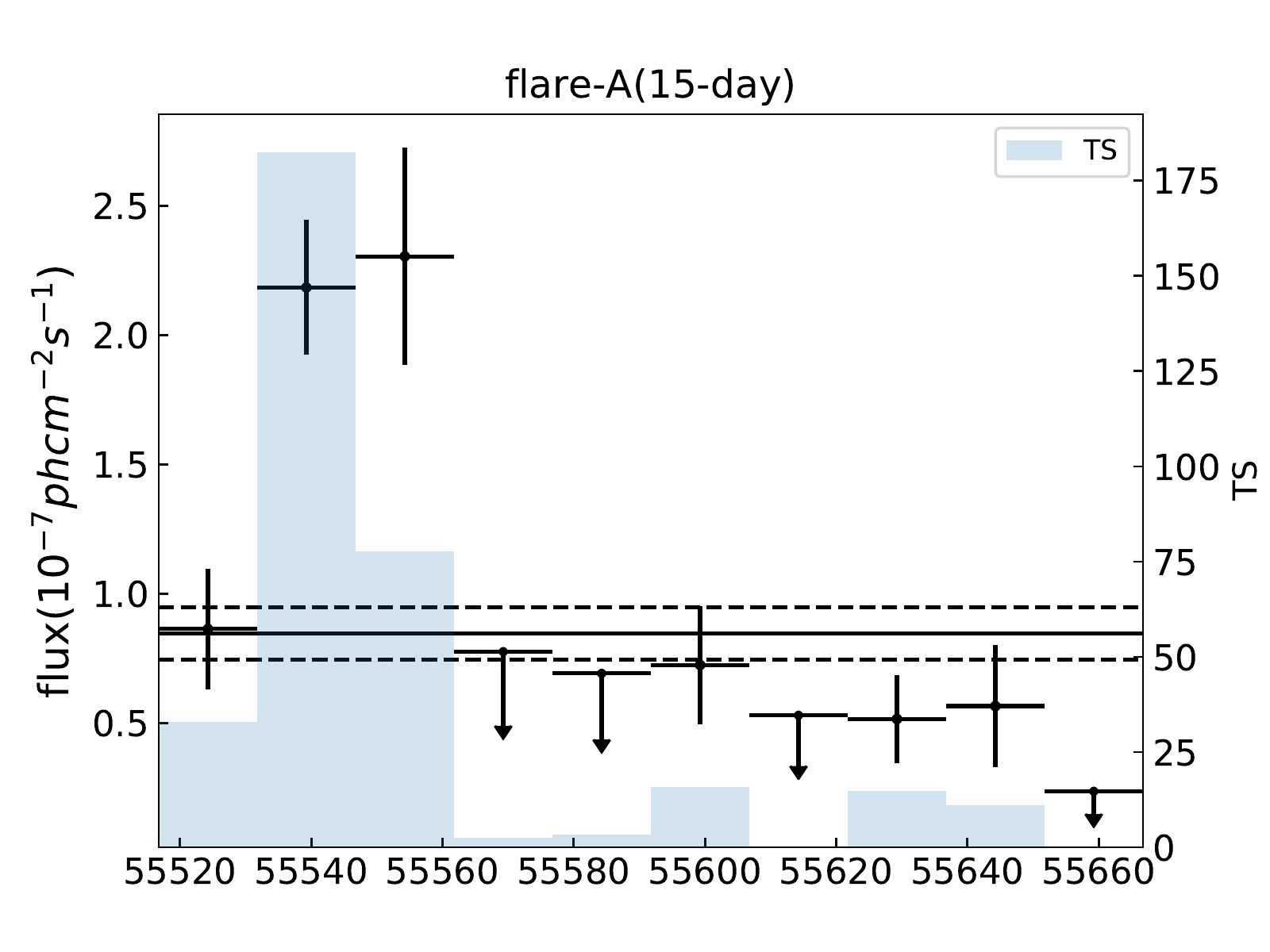}
\includegraphics[width=0.42\columnwidth]{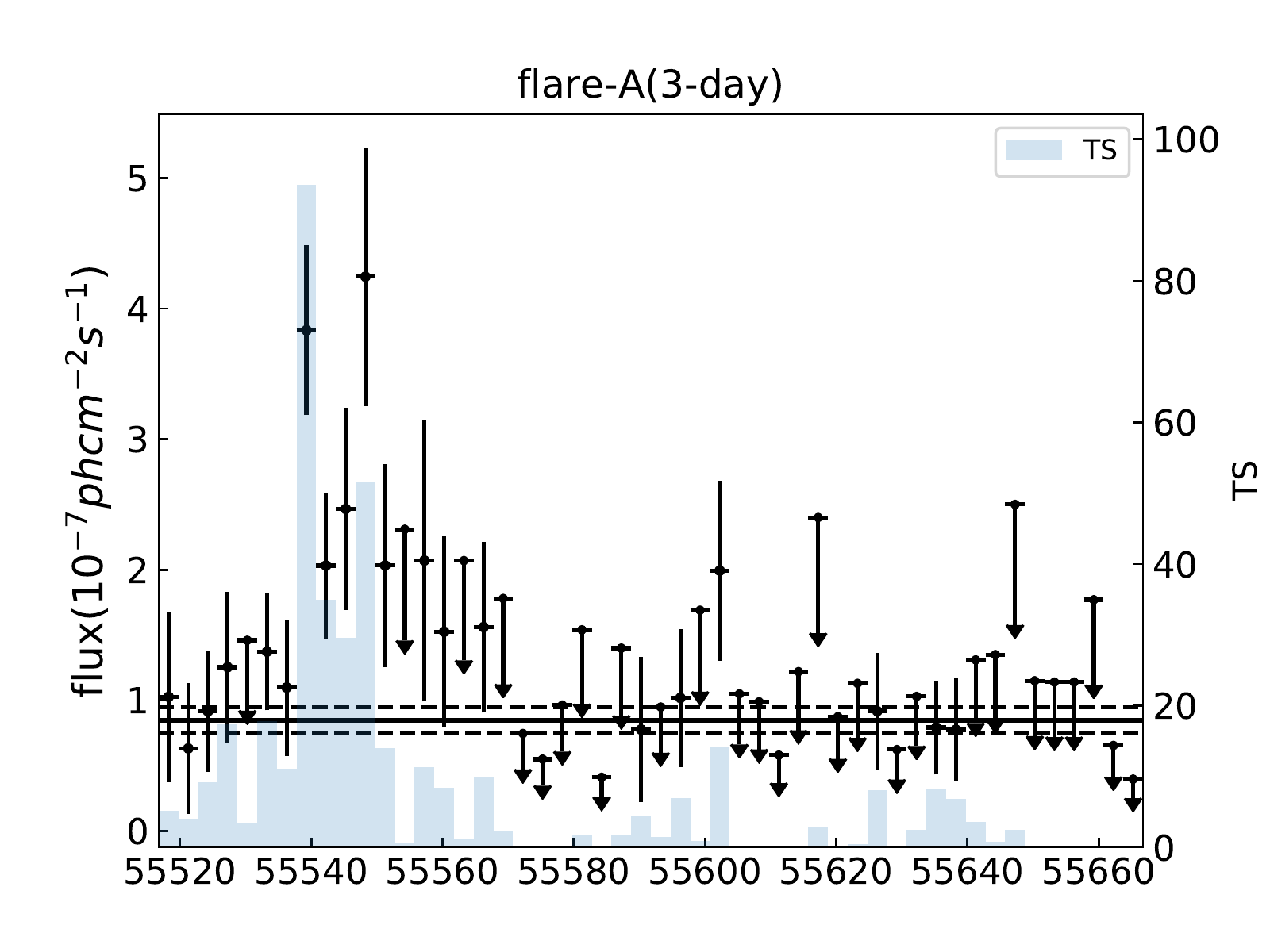}
\includegraphics[width=0.42\columnwidth]{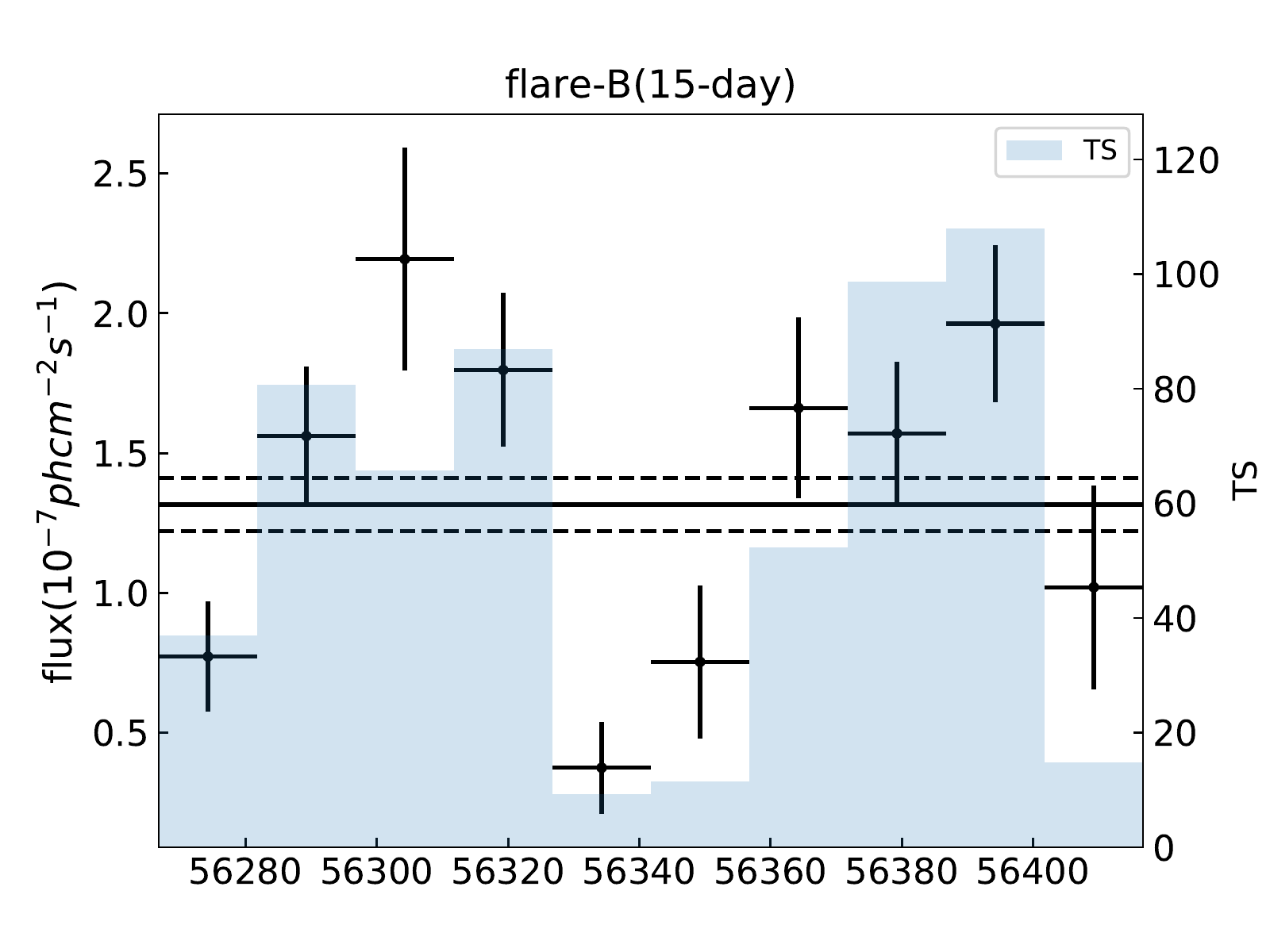}
\includegraphics[width=0.42\columnwidth]{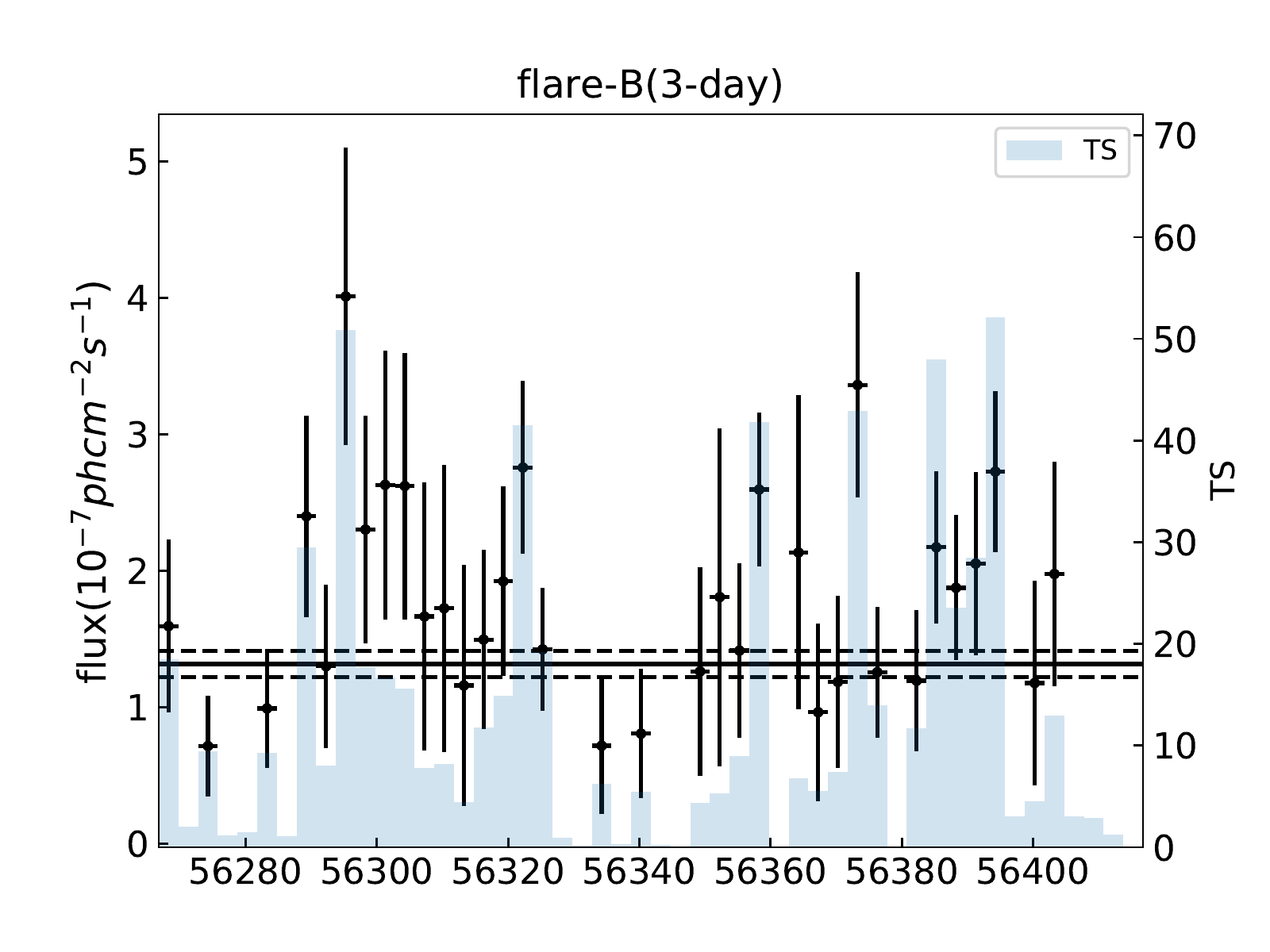}
\includegraphics[width=0.42\columnwidth]{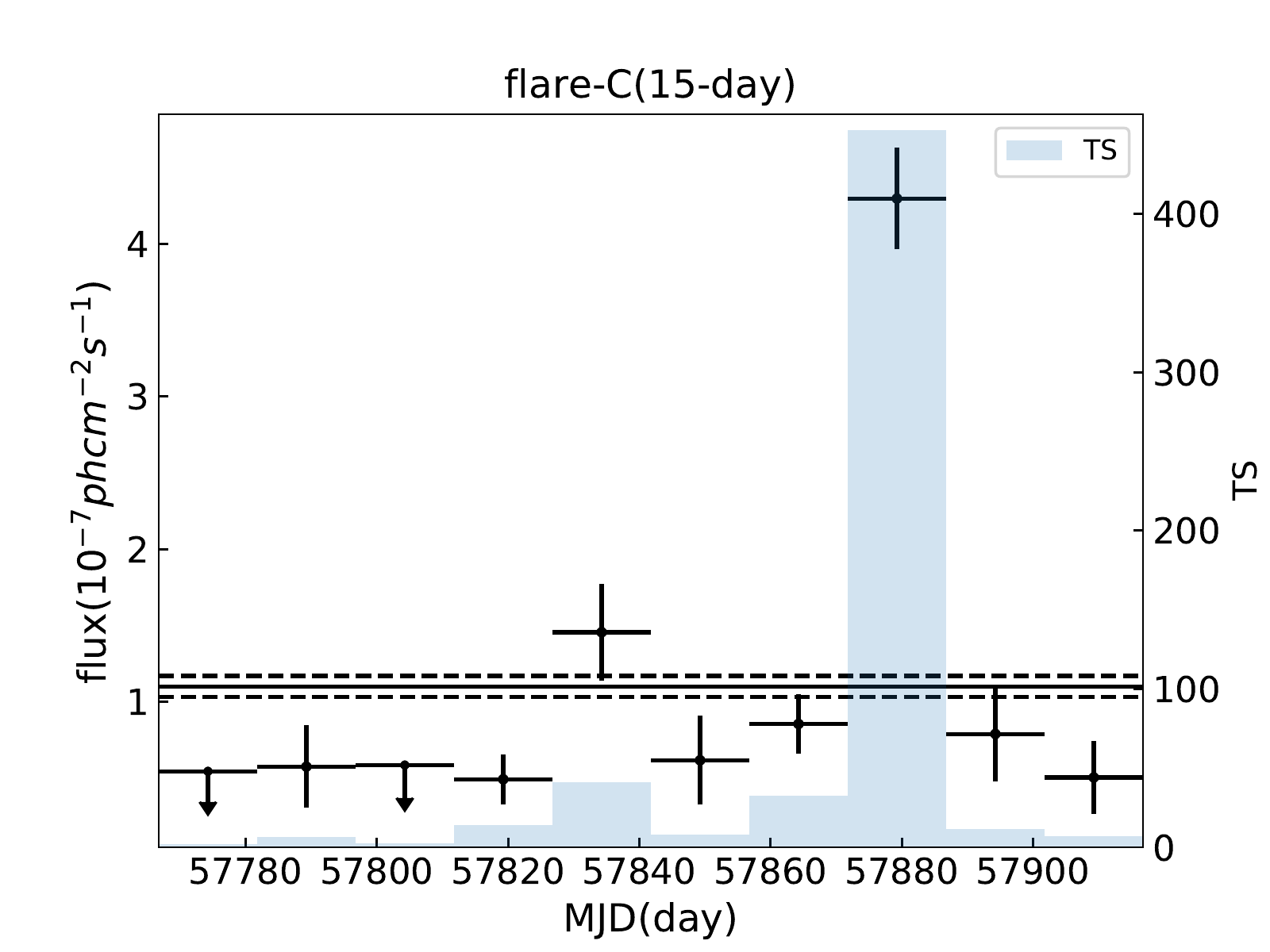}
\includegraphics[width=0.42\columnwidth]{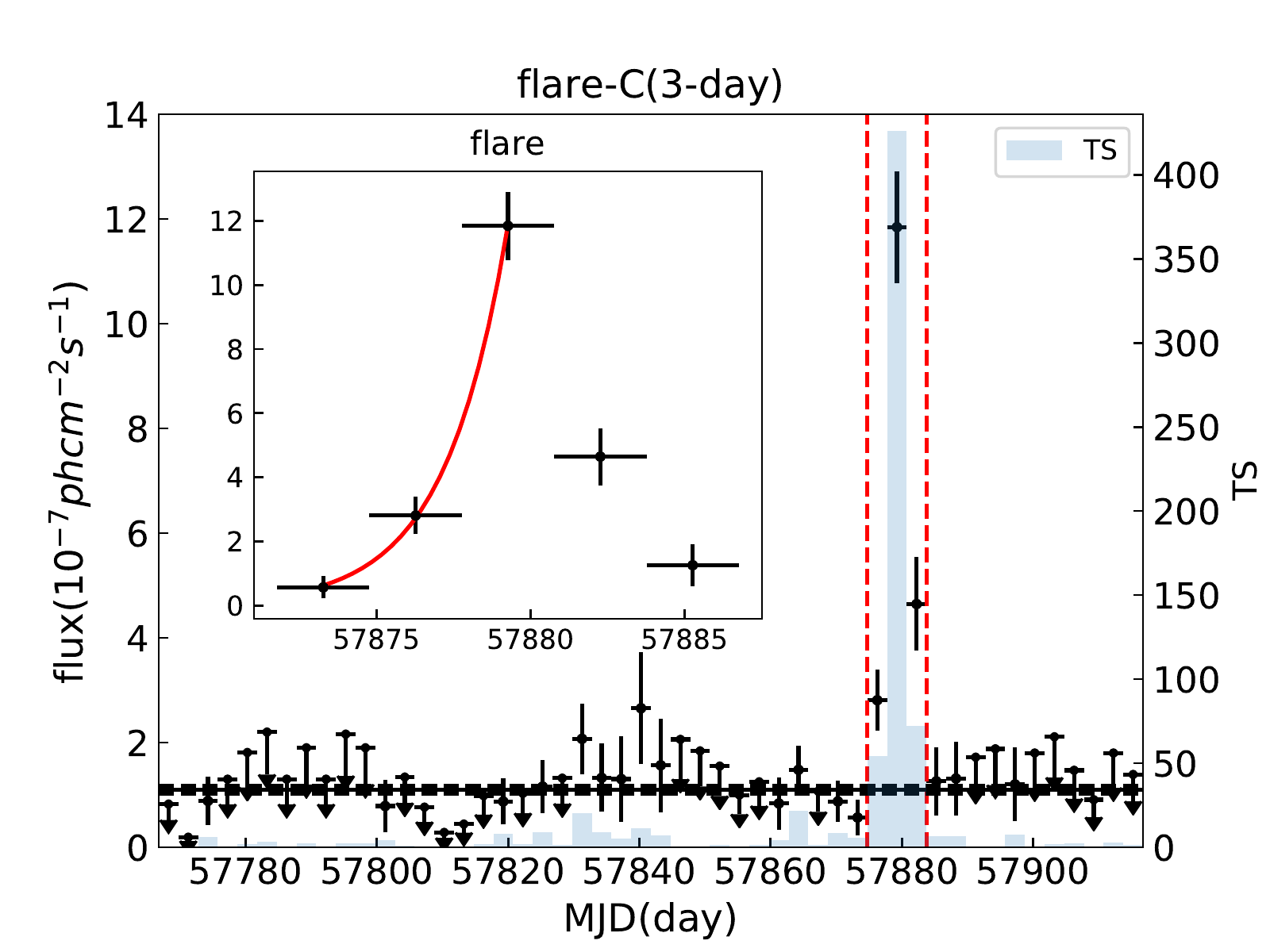}
\caption{15-day and 3-day time bin light curves correspond to the three flares of NVSS J053954$-$283956. Horizontal solid line along with two dashed lines in each panel represent the average flux and its $1\sigma$ error range then. In the 3-day time bin light curve of flare-C (i.e. the right bottom panel), red dashed vertical lines represent the time epoch when searches of intraday $\gamma$-ray variability are performed. Meanwhile, a zoomed-in panel of flare-C along with an exponential fit (red solid line) of the ascent phase is also presented here.}
\label{fig:lcb-15-3}
\end{figure}

\begin{figure}
\centering
\includegraphics[width=0.8\columnwidth]{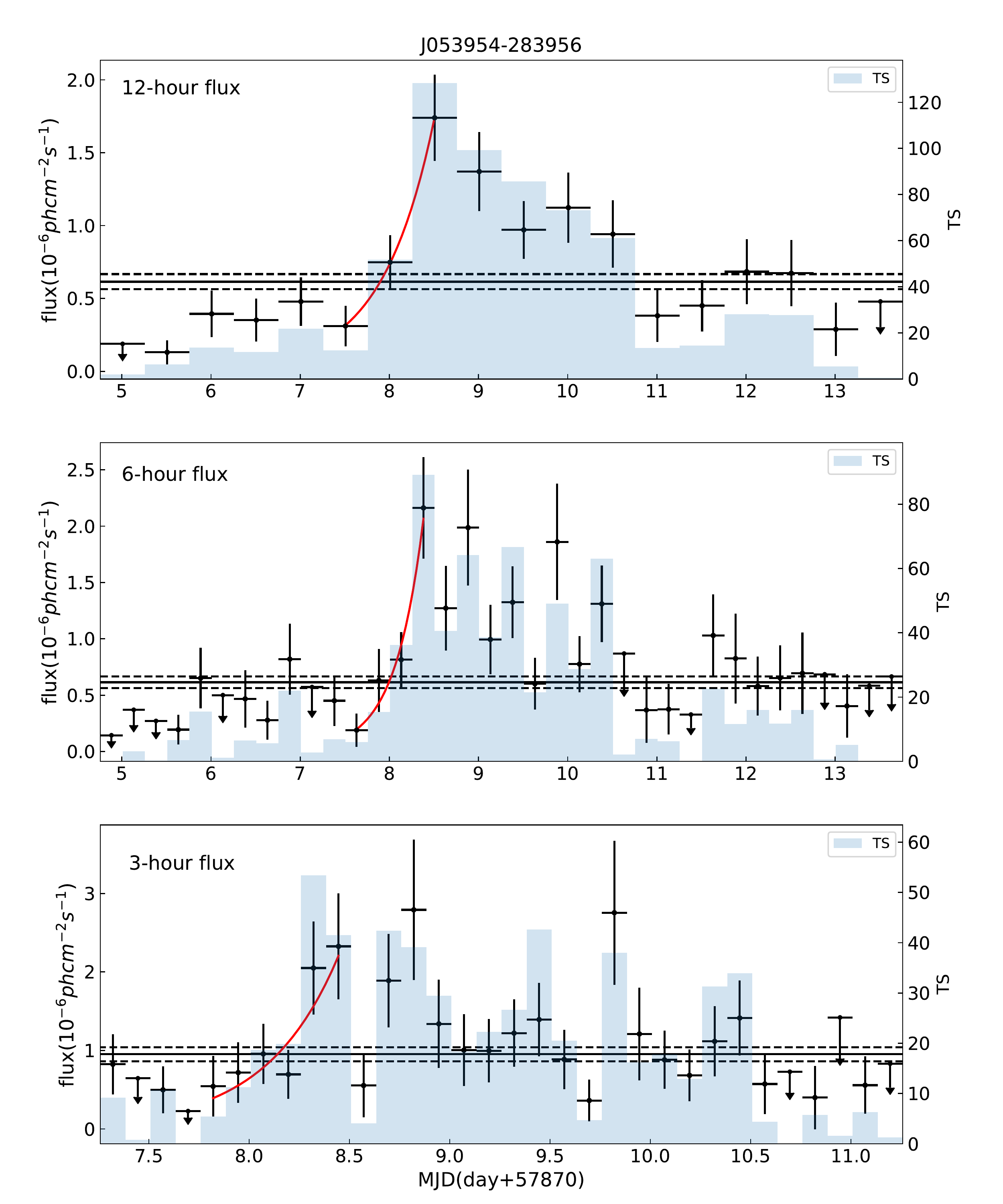}
\caption{12-hr (upper panel), 6-hr (middle panel) and 3-hr (bottom panel) light curves focusing on the flare-C of NVSS J053954$-$283956. Horizontal solid line as well as two dashed lines in each panel correspond to the average flux and its $1\sigma$ error range then. Red lines represent the exponential fits of the ascent phase of flare-C.}
\label{fig:lc6-3-3h}
\end{figure}

\subsubsubsection{NVSS J053954$-$283956}
NVSS J053954$-$283956, also named as PKS 0537$-$286, is one of the most luminous high-redshift quasars ($z$=3.104, \citealt{1978ApJ...226L..61W}). Its first detection was at radio frequencies \citep[e.g.,][]{1975AuJPA..34....1B}. It is also a bright and well-studied source in X rays \citep[e.g.,][]{1981ApJ...245..357Z,2007ApJ...669..884S,2010A&A...509A..69B}, showing an extremely high X-ray luminosity ($\rm \sim 10^{47}$ erg $\rm s^{-1}$ in the 0.1-2 keV range) and a particularly hard spectrum ($\rm \Gamma_{X}\sim$ 1.2), indicative of a significant contribution of the nonthermal jet emission. In temporal perspective, modest optical-NIR and X-ray continuum variations have been observed \citep{2010A&A...509A..69B}. Here we present its $\gamma$-ray temporal characteristics. As shown in the monthly $\gamma$-ray light curve of NVSS J053954$-$283956 (Figure \ref{fig:fiveLC}), there are three $\gamma$-ray flares. Enlarged 15-day and 3-day time bin $\gamma$-ray light curves corresponding to these flares are presented in Figure \ref{fig:lcb-15-3}. For the first two flares (i.e. flare-A and flare-B), no significant intraday $\gamma$-ray variability are found, confirming a previous study that suggests a minimum $\gamma$-ray variability timescale of $\sim$18 days then \citep{2013ApJ...767..103V}. However, the third flare (i.e. flare-C) detected in May 2017\footnote{This flare event has also been reported in ATel by {\it Fermi} collaboration \citep{2017ATel...10356...1C}.}, exhibits a totally different behavior. In the 15-day time bin light curve of the flare-C, flux of the eighth bin is significantly higher than the averaged flux, whereas the target maintains in the low flux state for the rest of time. Moreover, such a behavior is confirmed by the further 3-day time bin $\gamma$-ray light curve, where an intense $\gamma$-ray outburst suddenly appears. The flux quickly rises to the maximum value within about 6 days (1.46 days at the source frame) and the descent time is as short as the ascent time. In addition to the short variability timescale, variation amplitude of this outburst is one order of magnitude larger than the average flux. The 3-day peaking flux reaches to $\rm (1.2\pm0.1)\times 10^{-6}$ ph $\rm cm^{-2}$ $\rm s^{-1}$ with a very high TS value ($\simeq$ 426), while averaged fluxes in epoch of flare-C and entire 105 months are $\rm (1.0\pm0.1)\times 10^{-7}$ ph $\rm cm^{-2}$ $\rm s^{-1}$ and $\rm (4.9\pm0.2)\times 10^{-8}$ ph $\rm cm^{-2}$ $\rm s^{-1}$, respectively. These results put a tight constraint on the doubling timescale at the source frame, $\tau_{doub,source}=\Delta t\times {\rm ln}2/{\rm ln}(F_{1}/F_{2})/(1+z) < 0.5$ day. Powerlaw function provides an acceptable description to the burst SED while log-parabolic function does not bring any significant improvements. Spectral index of the burst SED ($\rm \Gamma$=2.53$\pm$0.09) is slightly harder than the averaged SED ($\rm \Gamma$=2.78$\pm$0.04), consistent with \cite{2017ATel...10356...1C}. Meanwhile, following $\gamma$-ray localization analysis gives that the optimized location at this time is R.A. 85.04$^{\circ}$ and DEC. -28.64$^{\circ}$. The corresponding 95\% C.L. error radius is 0.10$^{\circ}$ which is consistent with the value (0.08$^{\circ}$) listed in 3FGL \citep{2015ApJS..218...23A}. Since the angular separation between the $\gamma$-ray position and radio location of NVSS J053954$-$283956 is 0.06$^{\circ}$, it still locates within the 95\% C.L. error radius. Note that in the normal observation mode LAT performs a complete and uniform coverage of the sky in 3-hr, thus we limit the minimum time bin in our analysis to 3-hr. Therefore, 12-hr, 6-hr and 3-hr time bin $\gamma$-ray light curves are extracted to perform further investigations, see Figure \ref{fig:lc6-3-3h}. Since in the epoch of flare-C the target is at a high flux state that most of the time bins are not upper limits, we adopt a simple $\chi^{2}$ test to check whether the source is significantly variable. By optimizing the assuming constant flux, we find evidences of significant variability on intraday $\gamma$-ray light curves, ({\it p}, $\chi^{2}$/dof) = ($\rm 7.0\times 10^{-8}$, 61.4/14) for 12-hr light curve and ($\rm 1.6\times 10^{-5}$, 65.7/25) for 6-hr light curve, respectively. However, variability for 3-hr light curve is not statistically significant due to large uncertainties, ({\it p}, $\chi^{2}$/dof) = ($\rm 0.11$, 26.7/19). Variability timescales are also estimated (listed in Table \ref{table:tb2}) by fitting data in the ascent phase with the exponential function:
\begin{equation}
{F(t)=F({t}_{0})\cdot2^{-(t-{t}_{0})/\tau}},
\label{eq:time}
\end{equation}
where ${F(t)}$ and ${F({t}_{0})}$ are the fluxes at time ${t}$ and ${t}_{0}$, respectively, and $\tau$ is the characteristic time scale. In the 12-hr light curve, a quick raise begins in the sixth time bin (i.e. at MJD 57877.5, with a flux of $\rm (4.3\pm1.6)\times 10^{-7}$ ph $\rm cm^{-2}$ $\rm s^{-1}$), one day later the flux reaches the peak (i.e. at MJD 57878.5, $\rm (1.6\pm0.3)\times 10^{-6}$ ph $\rm cm^{-2}$ $\rm s^{-1}$). Since the eighth time bin, the target maintains in a high flux state ($\rm \gtrsim 10^{-6}$ ph $\rm cm^{-2}$ $\rm s^{-1}$) but with a relatively modest descent that costs about 2.5 days, then the target is back to a quiescent flux state. This variability trend is confirmed by the 6-hr light curve, from which the ascent time is constrained as 18-hr (as short as 4.4-hr at the source frame). Together with the variation amplitude (i.e. $\rm F_{bin15}/F_{bin12}\simeq$ 8.4), $\tau_{doub,source}$ can be estimated as short as 1.4-hr. A similar $\tau_{doub,source}$ (i.e. 1.3-hr) can be also derived from the ascent phase in 3-hr light curve despite the large error bars. Meanwhile, the 6-hr light curve reveals that the entire outburst may constitute of several sub-flares. Interestingly, violent variability may appear in these sub-flares. For example, in 3-hr light curve the flux raises from $\rm (3.8\pm2.7)\times 10^{-7}$ ph $\rm cm^{-2}$ $\rm s^{-1}$ at MJD 57879.70 to $\rm (2.5\pm0.9)\times 10^{-6}$ ph $\rm cm^{-2}$ $\rm s^{-1}$ at MJD 57879.82, leading to a very short variability timescale of $\tau_{doub,source}\sim$ 16-min.

\begin{figure}
\centering
\includegraphics[width=0.4\textwidth]{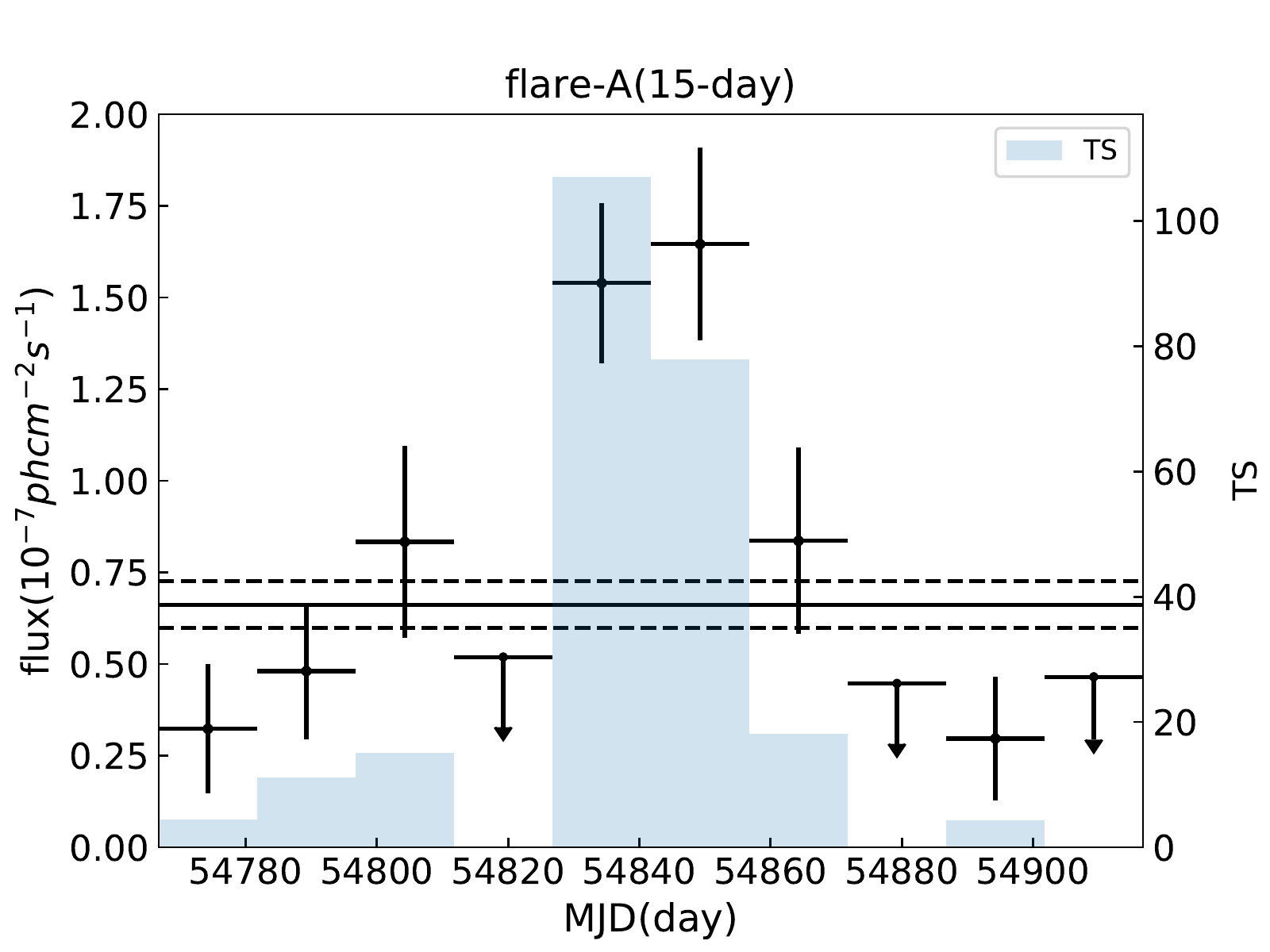}
\includegraphics[width=0.4\textwidth]{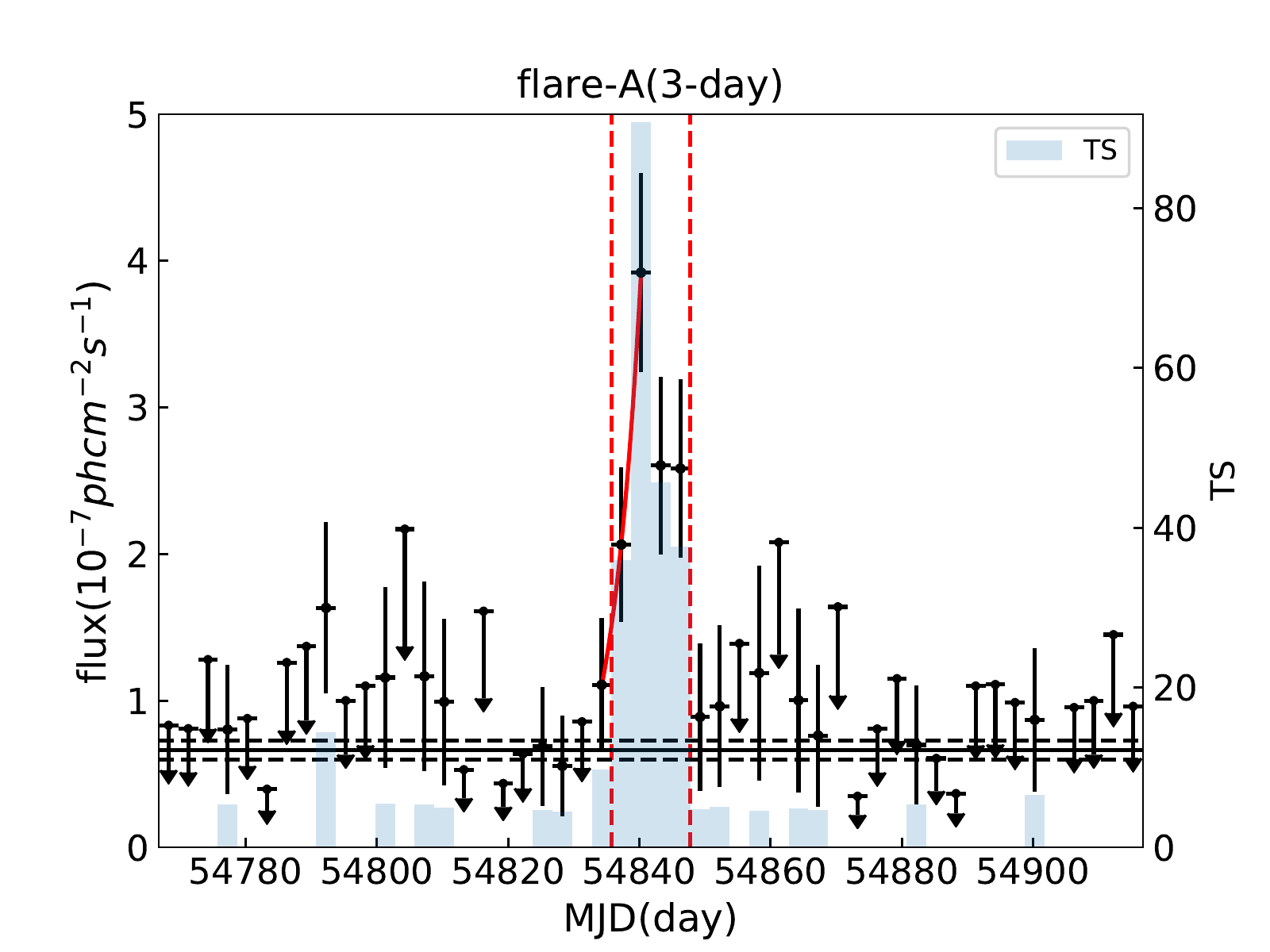}
\includegraphics[width=0.4\textwidth]{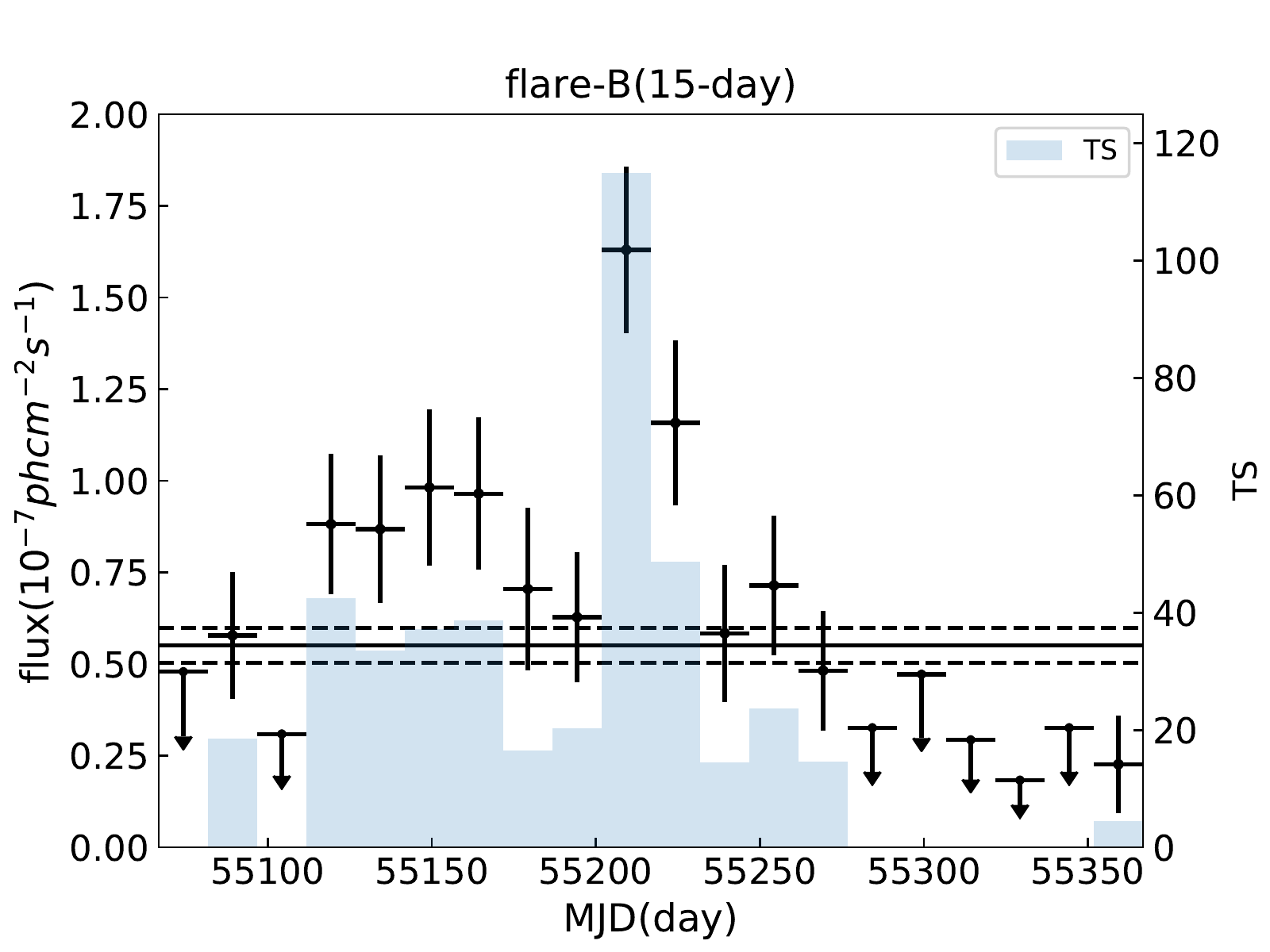}
\includegraphics[width=0.4\textwidth]{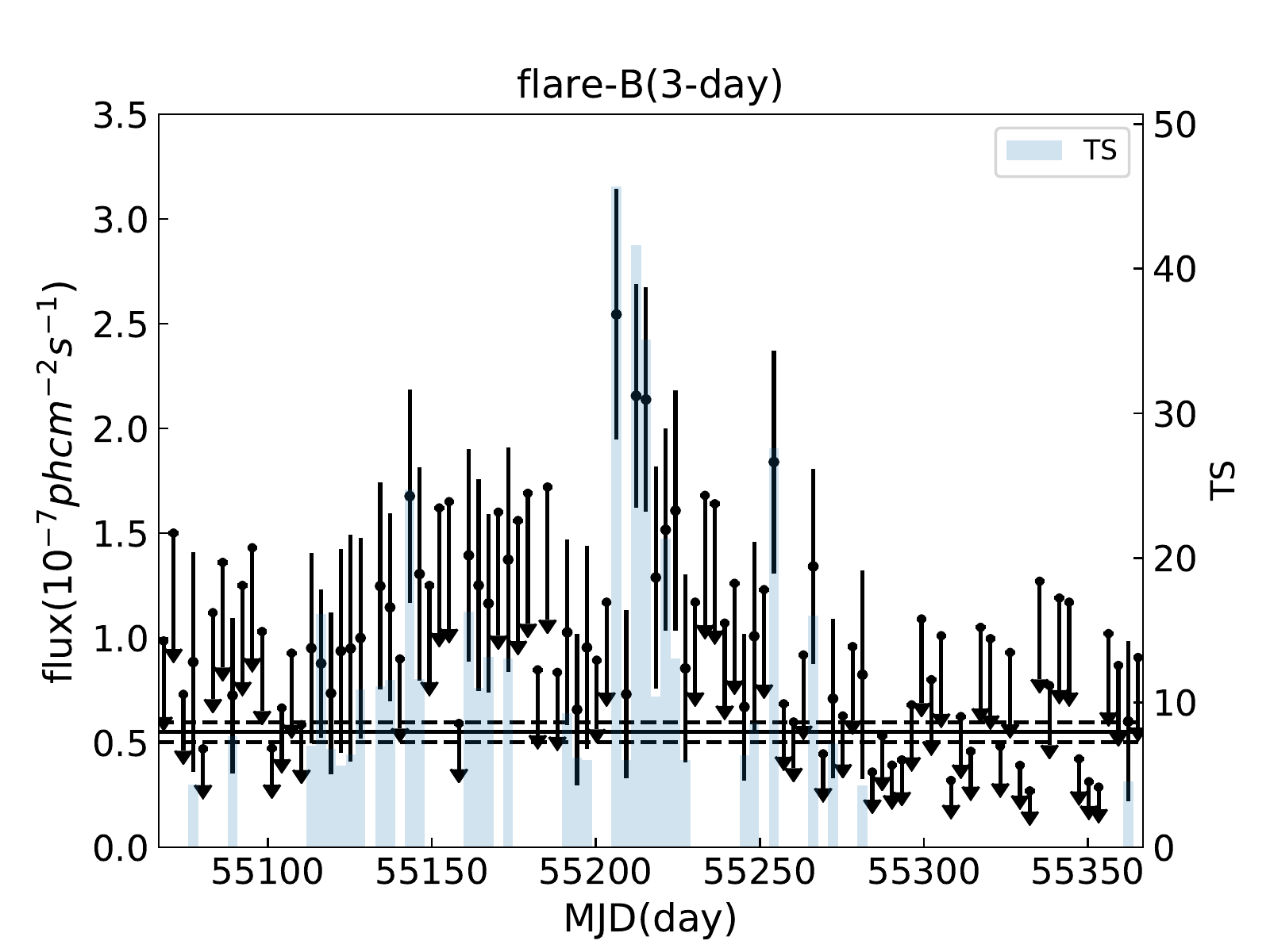}
\caption{15-day and 3-day time bin light curves correspond to the two flares of of NVSS J080518$+$614423. Horizontal solid line along with two dashed lines in each panel represent the average flux and its $1\sigma$ error range then. In the 3-day time bin light curve of flare-A (i.e. the right upper panel), red dashed vertical lines represent the time epoch when further temporal analyses are performed, along with an exponential fit (red solid line) of the ascent phase.}
\label{fig:lca-15-3}
\end{figure}

\begin{figure}
\centering
\includegraphics[width=0.4\textwidth]{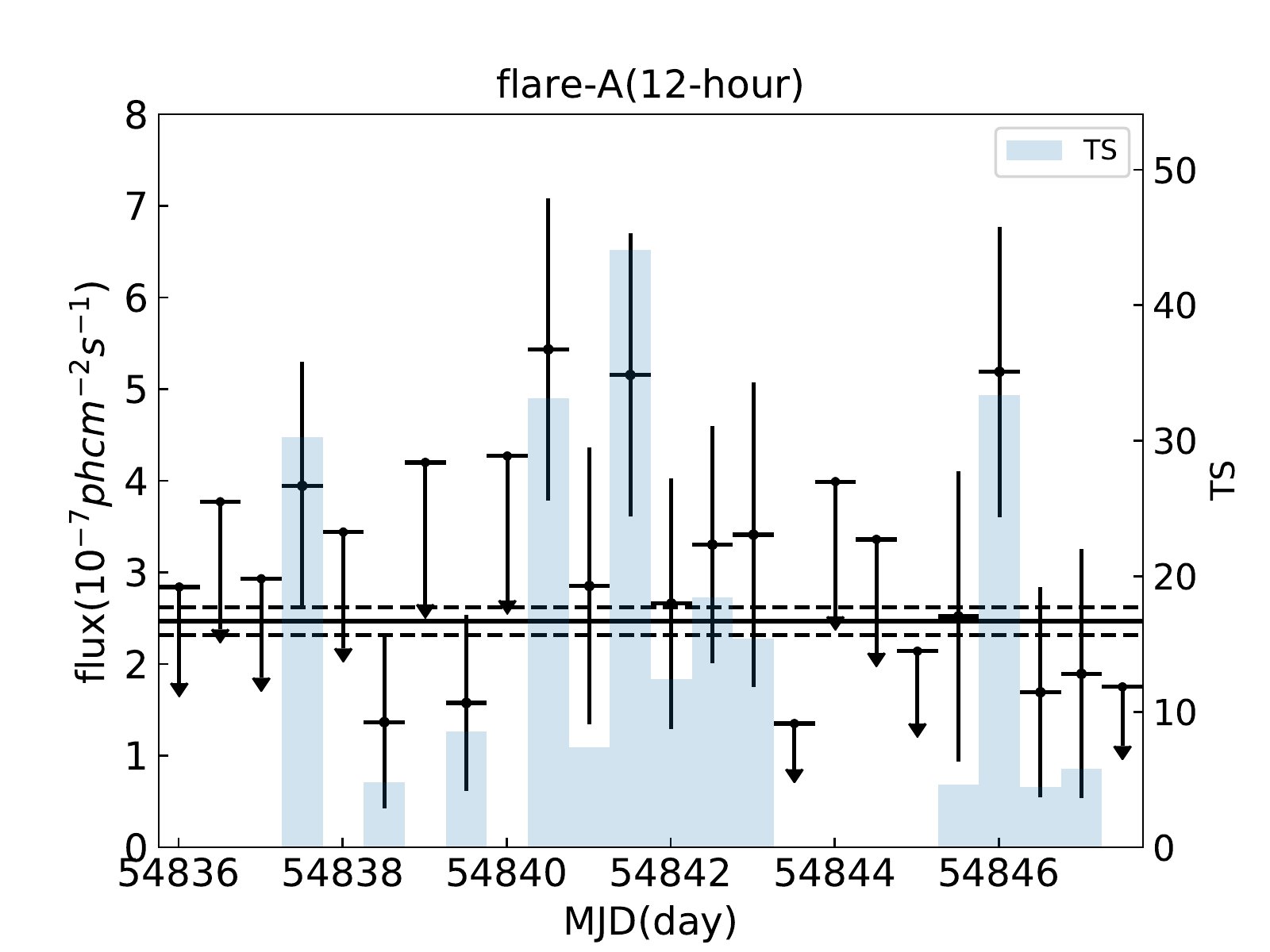}
\includegraphics[width=0.4\textwidth]{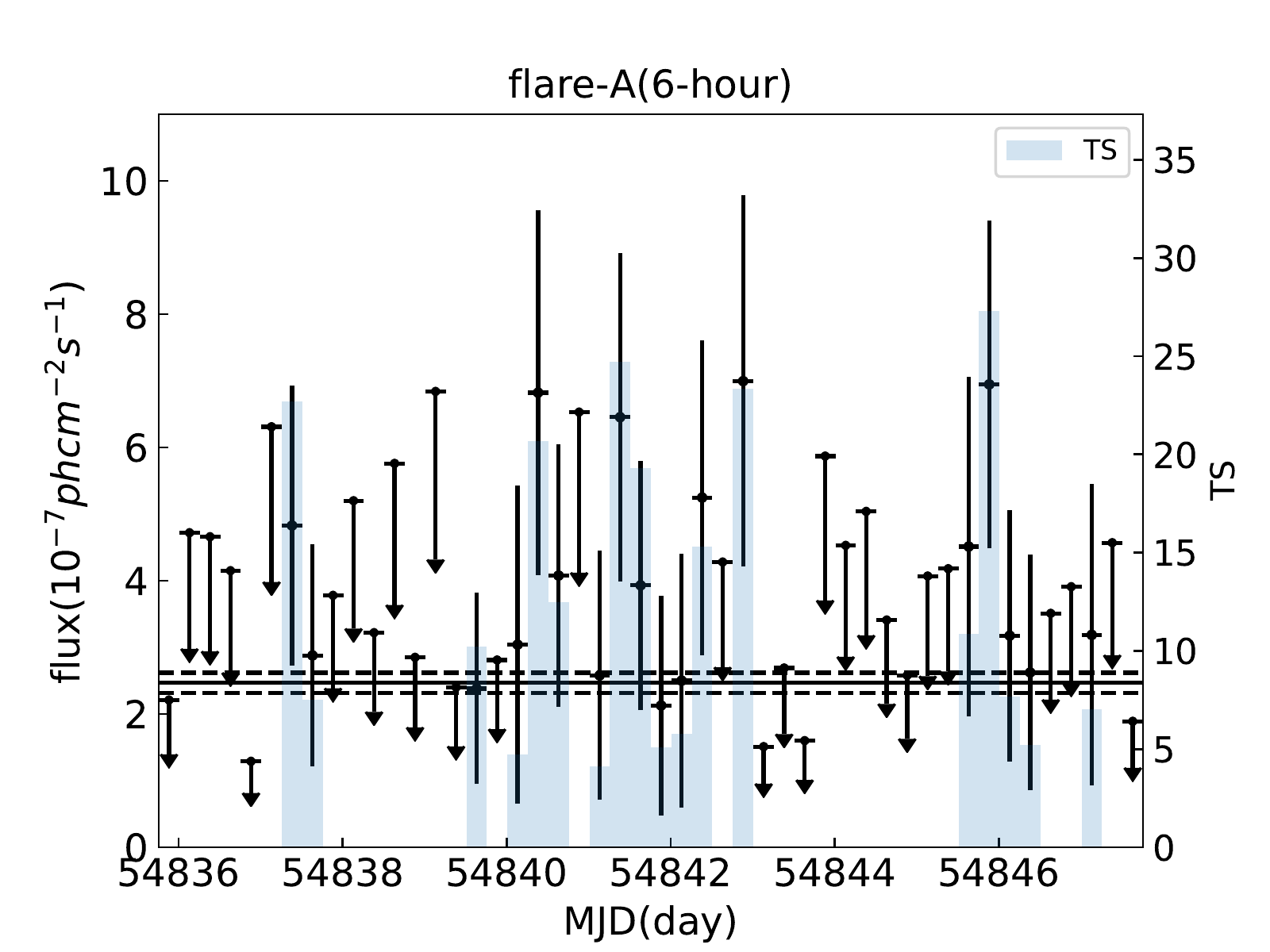}
\caption{12-hr (left panel) and 6-hr (right panel) light curves focusing on the flare-A epoch of NVSS J080518$+$614423. Horizontal solid line as well as two dashed lines in each panel correspond to the average flux and its $1\sigma$ error range then.}
\label{fig:lca-12-6}
\end{figure}

\subsubsubsection{NVSS J080518+614423}
NVSS J080518+614423 ($z$=3.033, \citealt{2005ApJ...626...95S}) was also firstly known as radio emitters \citep{1991ApJS...75....1B}, then its optical counterpart was identified \citep{2002MNRAS.329..700S}. Similar with NVSS J053954$-$283956, it has been detected by {\it WISE} and {\it Swift}-BAT \citep{2012ApJ...748...68D,2013ApJS..207...19B}. In its monthly $\gamma$-ray light curve, after two flares (i.e. flare-A and flare-B) in the first and second year of {\it Fermi} observation, it maintains at a low flux state for several years. These two flares are confirmed by enlarged 15-day time bin $\gamma$-ray light curves, see Figure \ref{fig:lca-15-3}. More importantly, an evidence of intraday $\gamma$-ray variability is found from the 3-day time bin $\gamma$-ray light curve of flare-A, of which the variability index is given as $\sigma_{var}$ = $10.8\sigma$. But no similar behavior can be found in flare-B due to large uncertainties, also see Figure \ref{fig:lca-15-3}. In consideration of the ascent time of 6 days, together with the variability amplitude of 3.5, the intrinsic doubling timescale in flare-A can be estimated as $\tau_{doub,source}\simeq$ 19.6-hr. Note that the 3-day peaking flux ($\rm (3.9\pm0.7)\times 10^{-7}$ ph $\rm cm^{-2}$ $\rm s^{-1}$) is roughly 15 times of the 105 months averaged flux. Intraday $\gamma$-ray light curves (i.e. 12-hr and 5-hr time bins) corresponding to this epoch have been also extracted, see Figure \ref{fig:lca-12-6}, no further constraints of the variability timescale are obtained. Similar with NVSS J053954$-$283956, the spectral index in flare-A of NVSS J080518+614423 ($\rm \Gamma$=2.40$\pm$0.12) is harder than the averaged SED ($\rm \Gamma$=2.82$\pm$0.05). Meanwhile, the optimized $\gamma$-ray position at this time is R.A. 121.25$^{\circ}$ and DEC. 61.73$^{\circ}$ with 95\% C.L. error radius of 0.12$^{\circ}$. The angular separation between the $\gamma$-ray position and radio position of NVSS J080518+614423 is only 0.04$^{\circ}$, supporting the association.

\begin{figure}
\centering
\includegraphics[width=0.8\columnwidth]{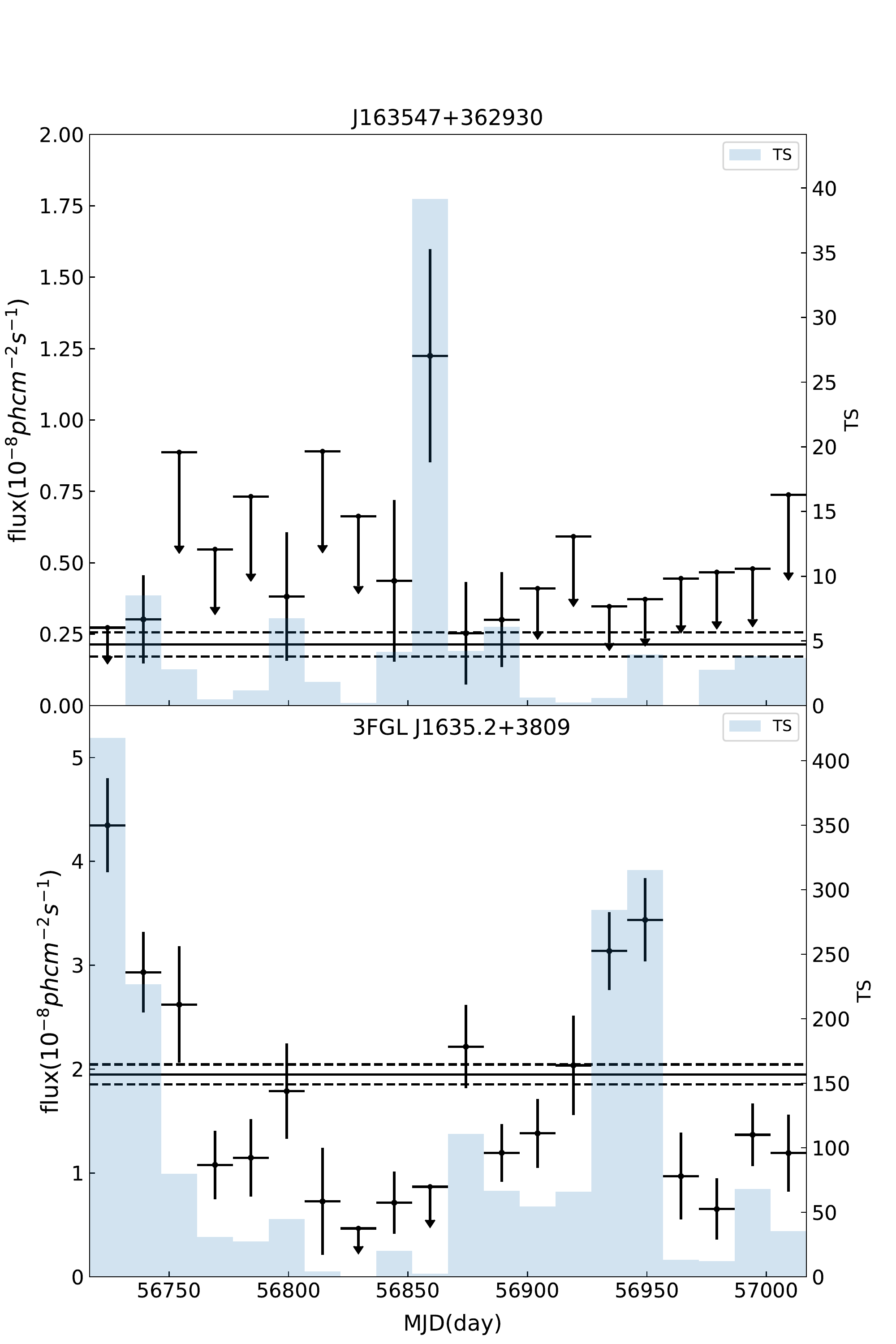}
\caption{500~MeV - 100~GeV 15-day time bin light curves of NVSS J163547+362930 as well as its neighbor 3FGL J1635.2$+$3809 focusing the flare-A epoch of the target. Horizontal solid line and two dashed lines in each panel represent the average flux and its $1\sigma$ error range of these two sources then.}
\label{fig:lc4-15}
\end{figure}

\subsubsubsection{NVSS J163547$+$362930}
NVSS J163547$+$362930 was discovered by the MIT-Green Bank 5~GHz survey \citep{1990ApJS...74..129G}. Then it has been included in the DR10 SDSS quasar catalog, with a redshift estimation of 3.647 \citep{2014A&A...563A..54P}. As shown in Figure \ref{fig:500mev}, there is a flare in the monthly $>$ 500~MeV $\gamma$-ray light curve of NVSS J163547$+$362930 which is probably not from the neighbor. Therefore, an enlarged 15-day time bin $>$ 500~MeV light curve has been extracted, together with another one of the neighbor 3FGL J1635.2+3809, see Figure \ref{fig:lc4-15}. Interestingly, though the variability at the overall period is not significant (i.e. $\sigma_{var}$ = $1.5\sigma$) due to large uncertainties, flux of the tenth bin (TS = 39) is roughly three times of the averaged flux during this epoch. While the neighbor is not detectable (TS $<$ 4) for {\it Fermi}-LAT at the same time, confirming what we find in the monthly light curve. The optimized $\gamma$-ray location for this time bin is R.A.= 248.89$^{\circ}$ and Dec.= 36.46$^{\circ}$. Since the angular separation is 0.06$^{\circ}$ while the 95\% C.L. $\gamma$-ray error radius of is 0.10$^{\circ}$, the association between NVSS J163547$+$362930 and the $\gamma$-ray source is confirmed. Similar with former two sources, a sign of bluer-when-brighter spectral variability behavior has been found for NVSS J163547$+$362930 (i.e. $\rm \Gamma_{flare,>500~MeV}$ = 2.58$\pm$0.25 while $\rm \Gamma_{105~month,>500~MeV}$ = 2.73$\pm$0.23). From the 15-day light curve, intrinsic doubling timescale can be constrained as $\sim$ 2.2 and 1.4 days for the ascend and descend phase, respectively. Unfortunately, no evidences of intraday $\gamma$-ray variability are found in further temporal analyses.

\section{SUMMARY AND DISCUSSIONS}
\label{sec4}
Benefited from large effective area and wide field of view of {\it Fermi}-LAT, together with its routine survey mode covering the entire sky in every 3-hr, our understanding of $\gamma$-ray variability properties of blazars has been profoundly improved. Significant $\gamma$-ray variability has been accepted as a common feature of $\gamma$-ray blazars, with variability amplitude up to more than two orders of magnitudes and variability timescale ranging from minutes to years \citep[e.g.,][]{2012ApJS..199...31N,2015NewA...34..134L,2016ApJS..226...17L,2016ApJ...824L..20A}. In long-term, evidences of quasi-periodic modulation of $\gamma$-ray emission of several blazars have been reported \citep[e.g.,][]{2015ApJ...813L..41A,2017ApJ...835..260Z}. In addition to flux variation alone, changes of $\gamma$-ray spectrum are often observed during different flux statuses \citep{2010ApJ...710.1271A}. Moreover, multi-wavelength campaign, including $\gamma$-ray observation as well as complementary observations from radio to X rays, becomes a regular approach to investigate the physical processes of AGN jet, and $\gamma$-ray emission of blazars is observed tightly connected with emissions in other widows of the electromagnetic radiation \citep[e.g.,][]{2008Natur.452..966M,2010Natur.463..919A,2014ApJ...783...83L,2016ApJS..226...17L}. In high redshift regime (i.e. $z>$2), $\gamma$-ray variability of several blazars has been detailedly studied, including 0836$+$710 ($z$=2.22; \citealt{2013A&A...556A..71A}, TXS 0536+135 ($z$=2.69; \citealt{2014MNRAS.444.3040O}), PKS 1830$-$211 ($z$=2.51; \citealt{2015ApJ...799..143A}), CGRaBS~J0225+1846 ($z$=2.69; \citealt{2016ApJ...825...74P}) and PKS 2149$-$306 ($z$=2.35; \citealt{2016MNRAS.455.1881D}). Among these sources, PKS 2149$-$306 and PKS 1830$-$211 are the most luminous ones with peaking $\gamma$-ray luminosities of 1.5$\times 10^{50}$ and 2.9$\times 10^{50}$ erg $\rm s^{-1}$ \citep{2016MNRAS.455.1881D,2015ApJ...799..143A}, respectively. Meanwhile, evidences of fast $\gamma$-ray variability have been claimed for 0836$+$710 and PKS 1830$-$211 ($\tau_{doub,source}\sim$ 2-hr; \citealt{2013A&A...556A..71A,2015ApJ...799..143A}). By comparison, blazars beyond redshifts 3 are focused in this study. On one hand, violent $\gamma$-ray variability with large amplitude (i.e. over one order of magnitude) for NVSS J053954$-$283956 and NVSS J080518$+$614423 is reported. Especially, the former has a peaking $\gamma$-ray luminosity of 1.1$\times 10^{50}$ erg $\rm s^{-1}$ (in this work we take a $\Lambda$CDM cosmology with $H_{0}=67~{\rm km~ s^{-1}~Mpc^{-1}}$, $\Omega_{\rm m}=0.32$, and $\Omega_{\Lambda}=0.68$; \citealt{2014A&A...571A..16P}), becoming the third most luminous source among the high redshift blazars and the most distant $\gamma$-ray flaring source so far. Meanwhile, the intrinsic doubling variability timescale of NVSS J053954$-$283956 is constrained as short as 1.4-hr, which makes it also the most distant source with known intraday $\gamma$-ray variability up to now. The energy dissipation mechanism corresponding to this extreme phenomenon could be magnetic reconnection process and ``minijets'' scenario \citep[e.g.,][]{2009MNRAS.395L..29G,2012ApJ...754L..33C,2015AAS...22521407B}. On the other hand, our study reveals significant long-term $\gamma$-ray variability in 3 of these sources. The significant $\gamma$-ray variability together with the results of $\gamma$-ray localization analyses in the flaring epoch, strongly support the blazar nature of these $\gamma$-ray sources. Meanwhile, the observed bluer-when-brighter spectral variability behaviors suggest that identification of flaring epochs of high redshift blazars is helpful for searching the distant high energy $\gamma$-ray photons and may be used to constrain extragalactic background light (EBL) models.

\begin{figure}
\centering
\includegraphics[width=0.8\columnwidth]{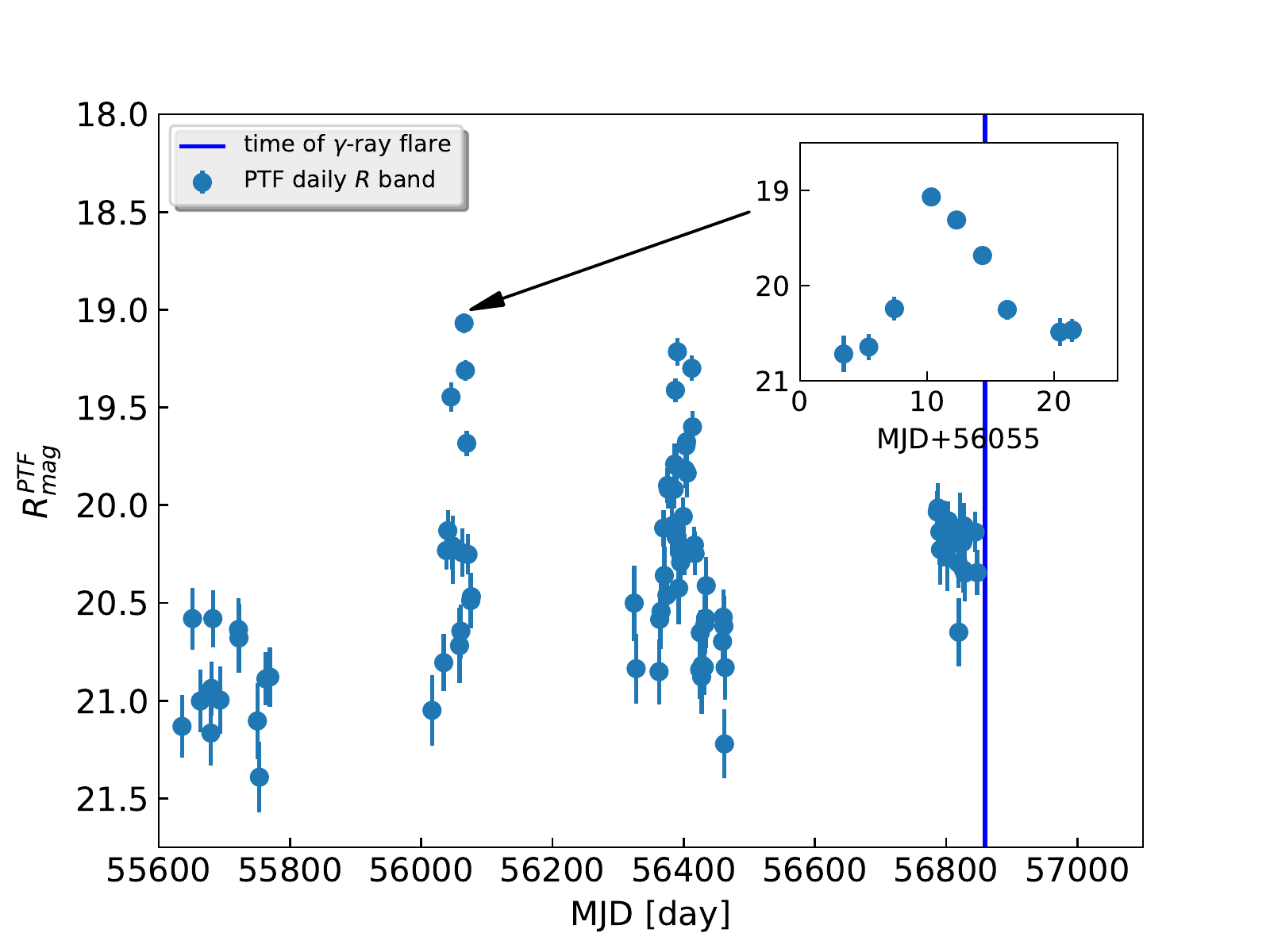}
\caption{Daily PTF/iPTF optical light curve of NVSS J163547+362930 with a zoomed-in panel of a flare peaking at MJD 55065.3.}
\label{fig:ptf}
\end{figure}

Since variation of optical emissions of FSRQs has been always observed simultaneously with variation of their $\gamma$-ray emissions, we also look up into archival Palomar Transient Factory (PTF)/intermediate PTF (iPTF) data to search addtional evidence of fast variability for these sources. The detailed description of the PTF/iPTF project can be found in \cite{2008SPIE.7014E..4YR}, and data reduction pipelines as well as photometric calibration procedures have been introduced in literature \citep[e.g.,][]{2009PASP..121.1395L,2012PASP..124..620,2012PASP..124..854O,2014PASP..126..674L,2015ASPC..495..197S}. Catalog Mould $R$ (i.e. $R_{\rm PTF}$) band {\tt SExtractor} \citep{1996A&AS..117..393B} data from the IPAC pipeline for all 7 high redshift blazars have been downloaded from the PTF/IPAC data archive hosted at the NASA/IPAC Infrared Science Archive (IRSA)\footnote{\url{http://irsa.ipac.caltech.edu/applications/ptf/}}. The $R_{\rm PTF}$ mag for each source is extracted by matching the detected sources in the catalogs to the input high redshift blazar with a match radius of two arc seconds. Unfortunately, except NVSS J151002$+$570243 and NVSS J163547$+$362930, the PTF/iPTF data of these sources are rather sparse (i.e. $\le$ 20 nights for 105 months). Meanwhile, the optical emission of NVSS J151002$+$570243 is faint (i.e. $r_{sdss}\simeq$ 20.3 mag), close to the detection limit of a routine 60 s $R_{\rm PTF}$ band exposure ($\sim$ 21 mag; \citealt{2012PASP..124..620}). Thus only light curve of NVSS J163547$+$362930 is presented. The light curve is extracted by a python script\footnote{An example of such a script can be found at \url{http://phares.caltech.edu/iptf/iptf\_SummerSchool\_2014/Miller2\_problems.html.}} with two steps. Firstly, several comparison stars (i.e. {\tt CLASS\_STAR} $>$ 0.95 and {\tt FLAGS} = 0) are picked under photometric condition (i.e. {\tt PHTCALFL = 1} and {\tt PCALRMSE}$<$ 0.04, 19 nights) based on their stable fluxes (i.e. $\rm mag_{std}<$ 0.04) throughout the entire time range. Then differential photometry based on these comparison stars is adopted to correct outliers affected by bad weather in ``raw'' light curve. As shown in Figure \ref{fig:ptf}, significant variability can be directly seen in the daily averaged PTF/iPTF light curve of NVSS J163547$+$362930 that includes 92 nights with time range between MJD 55635.5 and MJD 56847.2. There are four main optical flares in the light curve. Interestingly, in one flare that has well data coverage, the optical flux quickly rises from MJD 56060.4 with $R_{\rm PTF}$= 20.6 to MJD 56065.3 with $R_{\rm PTF}$= 19.1, indicating that the doubling time at the source frame then is $\sim$ 12-hr, see the zoomed-in panel of Figure \ref{fig:ptf}. This intraday optical variability behavior indicates that the central engine is highly active, consistent with the sign of fast $\gamma$-ray variability based on the 15-day $\gamma$-ray light curve, though no simultaneous iPTF observation is accessiable when the $\gamma$-ray flare appeares.

Based on the observed fast $\gamma$-ray variability, values of the Doppler factor of emitting jet blob should be high to avoid heavy absorption on $\gamma$ rays from soft photons via $\gamma\gamma$ process. The optical depth of $\gamma\gamma$ absorption can be calculated as \citep{1995MNRAS.273..583D}:
\begin{equation}
\tau_{\gamma\gamma}(x^{\prime})=\frac{\sigma_{\rm T}}{5}n^{\prime}(x^{\prime}_{\rm t})x^{\prime}_{\rm t}R^{\prime},
\end{equation}
where $\sigma_{\rm T}$ is the scattering Thomson cross section, $n^{\prime}(x^{\prime})$ is the number density of the target photon, $x^{\prime}_{\rm t}$ is the energy of the target photon in dimensionless units, and $R^{\prime}$ is the absorption length. The soft photons can be from the jet radiation itself and external emission from the accretion system (e.g. from accretion disk or broad emissions lines). Since $\gamma$ rays with energy $\simeq$ 3~GeV are detected during the flaring epoch of NVSS J053954$-$283956 and NVSS J080518$+$614423, energies of absorption soft photons could be at several keV and several tens of eV for the internal and external absorption, respectively. Only the former is considered here because few information of emissions at extreme ultraviolet wavelength is known. Adopting the variability timescale of 1.4-hr and 19.6-hr for these two sources and setting $\rm L_{soft}$ as $\rm 10^{47}$ erg $\rm s^{-1}$ \citep{2010MNRAS.405..387G}, we have $\delta \gtrsim$ 11 and $\delta \gtrsim$ 7. A similar calculation can be applied on NVSS J163547$+$362930 ($\delta \gtrsim$ 7 while $\rm E_{\gamma}\simeq$ 2~GeV) by assuming the optical and $\gamma$-ray emissions share a same radiation region. Therefore, radius of the radiation region (i.e. $R_{\gamma}<\delta c\tau_{doub,source}$) can be constrained as smaller than $\rm 1.7\times 10^{15}$ cm, $\rm 1.5\times 10^{16}$ cm and $\rm 9.1\times 10^{15}$ cm for NVSS J053954$-$283956, NVSS J080518$+$614423 and NVSS J163547$+$362930, respectively. The corresponding characteristic distance scale of the radiation region along the jet for a conical geometry is $r_{\gamma}\simeq R_{\gamma}/\theta \simeq R_{\gamma}\Gamma\simeq R_{\gamma}\delta$, where $\theta$ is the jet opening angle. Compared with the typical size of broad line region (i.e. $\sim$ 0.1 pc; \citealt{2010MNRAS.405L..94T}), the locations of $\gamma$-ray emission region of these three sources (i.e. $<$ 0.03 pc) could be within the broad line region (BLR). Meanwhile, the total jet power required to produce such high apparent $\gamma$-ray luminosities can be estimated as $\rm L_{jet} \simeq L_{\gamma}/(\eta_{jet}\Gamma^{2})$, where $\eta_{jet}$ is the jet radiative efficiency, typically $\sim$ 0.1 \citep{2012Sci...338.1445N}, and $\rm L_{\gamma}$ are $\sim$ $10^{50}$ erg $\rm s^{-1}$ and $10^{49}$ erg $\rm s^{-1}$ corresponding to NVSS J053954$-$283956 and NVSS J080518$+$614423, respectively. Since their Eddington luminosities are given as $\sim$ 3$\times 10^{47}$ erg $\rm s^{-1}$ and 2$\times 10^{47}$ erg $\rm s^{-1}$ \citep{2010MNRAS.405..387G}, their jet powers will exceed the Eddington luminositis if $\Gamma$ of the former source is smaller than 60 while the upper limit is 25 for the other source.

It is interesting to compare our results with those from SED modelings \citep{2010MNRAS.405..387G,2017ApJ...837L...5A}, though majority of the data used there are non-simultaneous. Doppler factor values of these high redshift blazars derived from the SED modelings are $\sim$ 11$-$15, consistent with our results. Meanwhile, similar with our results, locations of $\gamma$-ray emission region are also found within the BLR from SED modelings. Besides these approaches, direct measurements on the ejection speed of jet blob as well as the observed brightness temperature using the shortest radio variability timescale \citep[e.g.,][]{2009A&A...494..527H} and compared it with the theoretically expected brightness temperature assuming equipartition (i.e. $\rm T_{B}$ =  $\rm 5\times 10^{10}$ K; \citealt{1994ApJ...426...51R}), can also shed lights on the Doppler factor. However, these high redshift blazars are not included in current VLBA and radio flux monitoring projects \citep[e.g.,][]{2009AJ....137.3718L,2011ApJS..194...29R,2017ApJ...846...98J}. Nevertheless, let us make a comparison between our high redshift blazars and those at low redshifts. MOJAVE parsec-scale kinematics VLBA observations give an averaged apparent jet speed $\beta_{app}$ $\simeq 9$ of their sample \citep{2013AJ....146..120L}. Especially, observed $\beta_{app}$ of bright sources with detection of fast $\gamma$-ray variability are larger than 10 \citep{2005AJ....130.1418J}. Based on multiwavelength radio light curves, a mean value of the Doppler factor for F-GAMMA FSRQs is suggested as $\sim$12 \citep{2017MNRAS.466.4625L}. These indications from radio observations are confirmed by SED modeling studies \citep[e.g.,][]{2015MNRAS.448.1060G,2015ApJ...807...51Z}. Meanwhile, studies of the luminosity functions of blazars and
their parent populations allow for a constraint of the Doppler factor, which is in agreement with the kinematics radio observations \citep[e.g.,][]{2012ApJ...751..108A}. In conclusion, no significant difference of the Doppler factor between these 7 high redshift blazars (i.e. $z >$ 3) and ones nearby is found, and future simultaneous multiwavelength campaigns are helpful to further understand these highly active monsters.

According to the {\it Swift}-BAT detected high redshift blazars \citep{2009ApJ...699..603A}, the number of SDSS-FIRST detected jetted AGN with $z >$ 3 is fairly less than the expectation of the orientation based AGN unified scheme \citep{2011MNRAS.416..216V}. One possible explanation is that the averaged Lorentz factor of these {\it Swift}-BAT high redshift blazars (i.e. $\Gamma \sim$ 5) is generally lower than a routine value (i.e. $\Gamma \sim$ 15). Violent and rapid variability in blazars beyond redshift 3 are found in this study. Note that in addition to the two detections of FSRQs $\gamma$-ray variability with timescale of several minutes based on either IACT observation or ToO repointing observation from {\it Fermi}-LAT \citep{2011ApJ...730L...8A,2016ApJ...824L..20A}, there are only a few FSRQs whose minimum variability is $\sim$ 1-2 hrs (e.g., \citealt{2013ApJ...766L..11S,2013A&A...555A.138F,2015NewA...34..134L,2015ApJ...807...79H,2016ApJS..226...17L}). NVSS J053954$-$283956 is one of most violently active $\gamma$-ray FSRQs. Meanwhile, the light curves extracted here is under the survey mode operation of {\it Fermi}-LAT, due to the limited exposure time, the reported doubling timescales should be treated as upper limits only. Moreover, besides NVSS J053954$-$283956, intraday variations are detected in other two high redshift blazars (i.e. three in all seven), which suggests that such a phenomenon should not be rare for these sources. Therefore, this deficit could be due to selection effects rather than a gradient descent of averaged Lorentz factor of blazars at higher redshifts.

We appreciate helpful suggestions from the anonymous referee. This research has made use of data obtained from the High Energy Astrophysics Science Archive Research Center (HEASARC), provided by $\rm NASA^{\prime}$s Goddard Space Flight Center. This research has also made use of the NASA/IPAC Extragalactic Database and the NASA/IPAC Infrared Science Archive which are operated by the Jet Propulsion Laboratory, California Institute of Technology, under contract with the National Aeronautics and Space Administration. This research makes use of the SIMBAD database, operated at CDS, Strasbourg, France.

This work was supported in part by the National Basic Research Program of China (No. 2013CB837000), NSFC under grants 11525313 (i.e., Funds for Distinguished Young Scholars) and 11703093.

\begin{deluxetable}{lccccccccccccc}
%\scriptsize
\tabletypesize{\footnotesize}
\tablenum{1}
\tablewidth{0pt}
\tablecaption{Basic information and $\gamma$-ray properties of the known 7 $\gamma$-ray blazars beyond redshift 3.}
\tablehead{
\colhead{NVSS name} &\colhead{{\it b}} &\colhead{{\it z}} &\colhead{$\rm F_{1.4~GHz}$} &\colhead{$ R_{mag}$} &\colhead{$M_{BH,\odot}$} &\colhead{$\rm F_{\gamma}$} &\colhead{$\Gamma_{\gamma}$} &\colhead{TS}  &\colhead{$\rm \sigma_{var}$}}
\startdata

J053954$-$283956 &$-27.3^{\circ}$ & 3.1 & 0.86 & 19.0 &9.3 &$4.92\pm0.23$ &$2.80\pm0.04$ & 1737.8  & 15.7 \\[3pt]
J064632$+$445116 &$17.5^{\circ}$ & 3.4 &0.45 & 18.5 &9.1  &$1.35\pm0.19$ &$2.95\pm0.13$ & 103.3 & 1.2 \\[3pt]
J080518$+$614423 &$32.4^{\circ}$ & 3.0 & 0.83 &19.6 &9.07 &$2.21\pm0.13$ &$2.83\pm0.05$ & 467.2  & 10.8 \\[3pt]
J135406$-$020603 &$50.1^{\circ}$ & 3.7 & 0.73 & 19.2 &8.9 &$1.0\pm0.16$ &$2.83\pm0.13$ & 60.1  & 5.0 \\[3pt]
J151002$+$570243 &$50.3^{\circ}$ & 4.3 & 0.20 & 19.9 &8.5  &$0.42\pm0.13$ &$2.72\pm0.20$ & 32.7 & 3.2 \\[3pt]
J163547$+$362930 &$42.1^{\circ}$ & 3.6 & 0.15 & 20.6 &8.7 &$1.93\pm0.29$ &$3.21\pm0.12$ & 129.9  & 6.2 \\[3pt]
J212912$-$153841 &$-41.9^{\circ}$ & 3.3 & 0.59 &16.5 &9.8  &$1.62\pm0.19$ &$3.04\pm0.12$ & 108.9  & 2.4 \\[3pt]

\enddata
\tablecomments{(1) NVSS name of the object; (2) Galactic latitude; (3) redshift; (4) NVSS radio flux density at 1.4~GHz in Jy; (5) apparent $R$ band magnitude; (6) logarithm of black hole mass; (7) average $\gamma$-ray flux of 105 months LAT data analysis in scale of $\rm 10^{-8}$ ph $\rm cm^{-2}$ $\rm s^{-1}$; (8) $\gamma$-ray spectral index corresponding to col. (7); (9) TS value corresponding to col. (7); (10) significance of the variability estimated from the monthly $\gamma$-ray light curve. The Galactic latitudes are derived from NED. The redshifts and the {\it R} magnitudes are obtained from Half Million Quasars catalog \citep{2015PASA...32...10F}. The NVSS 1.4~GHz flux densities are from \cite{1998AJ....115.1693C}. Note that the $\rm \sigma_{var}$ here for NVSS J163547$+$362930 could be influnced by the neigbor 3FGL J1635.2$+$3809. More details see Section \ref{sec:longterm}.}
\label{table:tb1}
\end{deluxetable}

\begin{deluxetable}{ccccc}
%\scriptsize
\tablenum{2}
\tablewidth{0pt}
\tablecaption{Flux doubling timescales}
\tablehead{
\colhead{NVSS Name} &\colhead{Epoch} & \colhead{Time bin} & \colhead{$\tau_{doub,source}$ (hour)} & \colhead{p-value}}
\startdata
J053954$-$283956 &flare-C &   3-day   & $8.3\pm0.2$ & ${2.9{\times}10^{-50}}$  \\[3pt]
J053954$-$283956 &flare-C &   12-hour & $2.4\pm0.6$ & ${7.0{\times}10^{-8}}$ \\[3pt]
J053954$-$283956 &flare-C &   6-hour  & $1.3\pm0.3$ &  ${1.6{\times}10^{-5}}$ \\[3pt]
J053954$-$283956 &flare-C &   3-hour  & $1.5\pm0.6$ & 0.11 \\[3pt]
J080518$+$614423 &flare-A &   3-day   & $19.6\pm2.5$ & ${1.0{\times}10^{-27}}$ \\[3pt]
\enddata
\tablecomments{(1)NVSS name of the object; (2) Data time epoch given in Figure \ref{fig:fiveLC}; (3) Estimated doubling timescales at the source frame along with 1$\sigma$ errors; (4) Probability that the null hypotheis stands.}
\label{table:tb2}
\end{deluxetable}

\end{document}